\documentclass[useAMS,usenatbib]{mn2e}

\usepackage{graphicx}  
\usepackage{amsmath}
\usepackage{soul}
\usepackage{color}
\def\aj{AJ}
\def\apj{ApJ}
\def\apjl{ApJ}
\def\apjs{ApJS}
\def\aap{A\&A}
\def\aaps{A\&AS}
\def\aapr{A\&AR}
\def\mnras{MNRAS}
\def\nat{Nature}
\def\pasp{PASP}%
\def\araa{ARA\&A}

\title[Stellar populations of Virgo dwarf ellipticals]{Stellar
  populations of Virgo cluster early-type dwarf galaxies with and
  without discs: a dichotomy in age?\thanks{Based on observations collected at the European
    Organisation for Astronomical Research in the Southern Hemisphere,
    Chile (programme 078.B-0178)}}
\author[S. Paudel et al.]{Sanjaya Paudel$^{1}$\thanks{E-mail:
    sjy@x-astro.net}, Thorsten Lisker$^{1}$, Harald
  Kuntschner$^{2}$, Eva K. Grebel$^{1}$,\newauthor Katharina Glatt$^{1}$\\
 $^{1}$Astronomisches Rechen-Institut, Zentrum f\"ur Astronomie der
  Universit\"at Heidelberg, M\"onchhofstr.\ 12-14, 69120
  Heidelberg, Germany\\
 $^{2}$Space Telescope European Coordinating Facility, European
 Southern Observatory, Karl-Schwarzchild-Str. 2, 85748 Garching, Germany}
\begin{document}
\date{  Accepted ...  Received ... ; in original form \today }

\pagerange{\pageref{firstpage}--\pageref{lastpage}} \pubyear{2009}

\maketitle

\label{firstpage}
\begin{abstract}
Using VLT/FORS2 spectroscopy, we have studied the properties of the
central (inner 320 pc) stellar populations of a sample of 26 nucleated early-type dwarf
(dE) galaxies in the Virgo Cluster of magnitude range: -18.59 $\leqslant$ M$_{\emph{r}}$ $\leqslant$ -15.39 mag. With the addition of the
data of Michielsen et al. (2008), extending the sample to 38
dEs, we find that these galaxies do not exhibit the same average stellar
population characteristics for different morphological subclasses. The
nucleated galaxies without discs, which are, on average, fainter than
dEs with discs and distributed in regions of higher local
density, are older and more metal poor (mean ages 
7.5 $\pm1.89$ Gyr and 3.1 $\pm0.83$ Gyr, mean
[Z/H] = $-$0.54 $\pm$0.14 dex and $-$0.31 $\pm$0.10 dex,
respectively).
 The
$\alpha$-element abundance ratio appears consistent with the solar value
for both morphological types.
Besides a well-defined relation of metallicity and luminosity, we also
find a clear anti-correlation between age and luminosity. More
specifically, there appears to be a bimodality: brighter galaxies
($M_r\leq -16.5$ mag), including the discy ones, exhibit
significantly younger ages than fainter dEs ($M_r\geq -17.0$
mag). The magnitude overlap between these two subgroups appears to be
resolved when considering in addition the weak correlation between age
and local density, such that older galaxies \emph{at a given
  magnitude} are located at higher densities. Thus there seems to be
no significant difference between the stellar
populations of dEs with and without discs \emph{when compared at the
  same magnitude and density}. We thereby revisit the question
whether both could belong to the same intrinsic population, with discs
surviving only in lower-density regions. However, it appears less
likely that fainter and brighter dEs have experienced the same evolutionary
history, as the well-established trend of
decreasing average stellar age when going from the most luminous ellipticals towards
low-luminosity Es and bright dEs is broken here.
The older and more metal-poor dEs could have had an early termination
of star formation activity, possibly being ``primordial'' galaxies in
the sense that they have formed along with the protocluster or
experienced very early infall. By contrast, the younger and relatively
metal-rich brighter dEs, most of which have discs, might have
undergone structural transformation of infalling disc galaxies. 
\end{abstract}


\section{Introduction}

Dwarf elliptical (dE) galaxies are the numerically dominant population in
the present-day Universe (\citealt{Sandage85}; \citealt{Binggeli87}). They
 also exhibit strong clustering, being found predominantly in the
close vicinity of giant galaxies, either as satellites of
individual giants, or as members of galaxy clusters (\citealt{Ferguson89};
\citealt{Ferguson94}). Unlike classical elliptical galaxies,
stellar population studies show that these galaxies exhibit on average
younger ages as compared to their giant
counterparts, and also a lower metal content according to the
correlation of metallicity and luminosity \citep[hereafter
  M08]{Michielsen08}. More detailed studies of dEs and 
the fainter dSphs in the Local Group, based on resolved stellar
populations, reveal that most dE/dSphs have a fairly extended star
formation history (e.g., \citealt{Mateo98}; \citealt{Grebel04}), with
the last star formation activity ranging from a few gigayears ago to
the time of re-ionisation. A number of studies in different clusters
(e.g., \citealt{Poggianti01}; \citealt{Rakos01}; \citealt{Caldwell03};
\citealt{Geha03}; \citealt{van04}) also provide a wide range of ages
for passive dwarfs. Another study, however, based on the velocity
distribution of the Virgo dEs (\citealt{Conselice01}), argue that dEs
are not an old cluster population. This evidence suggests that dEs in 
nearby clusters have a mixed origin. Some have properties consistent
with an early ``primordial'' formation, while others appear to be more
recently formed, possibly from the transformation of progenitors with
different morphological types.

Due to the distance of the Virgo cluster (16.5 Mpc, \citealt{Mei07}),
it is basically impossible to perform resolved stellar population
studies of dEs in Virgo (see, however, the studies of the red giant
branch by \citealt{Caldwell06}).
 Therefore, in order to probe
the star formation history of these faint galaxies, we use
integrated spectra with a good signal-to-noise ratio (SNR), providing a
sensitive tool for stellar population studies through absorption
feature measurements. The basic idea, which was
first introduced by \citealt{Burstein84}, is to provide a set of optical
absorption line strength indices, commonly known as the Lick/IDS system or
Lick indices (\citealt{Burstein84}; \citealt{Worthey94a};
\citealt{Trager98}). It is now a widely used approach to determine the
age, metallicity, and $\alpha-$element abundance of galaxies from
integrated spectra. The measured indices
from the optical spectrum of a galaxy are compared to their values
predicted by stellar population models, which are provided for example
by \citealt{Worthey94b}; \citealt{Vazdekis99}; \citealt{Bruzual03};
\citealt{Thomas03}; \citealt{Schiavon07}). The  SSP-equivalent age and metallicity
of a given galaxy are
then determined as the ones of the model whose indices best agree
with the measured ones.

It has  already been noticed that dwarf elliptical galaxies of the Virgo
cluster are consistent with solar [$\alpha$/Fe]-abundance ratios
\citep{Gorgas97}, indicating the slow chemical evolution in low-mass
systems. Other studies (\citealt{Geha03};
\citealt{van04}) confirmed this result.

As the most numerous type of galaxy in clusters, early-type dwarf
galaxies are ideal probes to study the physical processes that govern
galaxy formation and evolution in environments of different
density. The pronounced morphology-density relation
(e.g., \citealt{Dressler80}; \citealt{Binggeli87}) suggests that
early-type dwarfs were either formed mainly in high-density
environments, or originate from galaxies that fell into a cluster and
were morphologically transformed. Because having low masses their
properties are expected to depend strongly on the environment they
reside in. However, the actual formation
mechanisms are still a matter of debate (see \citealt{Jerjen05}, and
references therein). Most of the proposed scenarios are based on the
vigorous forces acting within a cluster environment, such as
interactions with a hot intracluster medium via pressure confinement,
shocks or ram pressure stripping (\citealt{Gunn72}; \citealt{Faber83};
\citealt{Silk87}; \citealt{Babul92}; \citealt{Murakami99}) and are able to
transform dwarf irregular (dIrr) galaxies or late-type spirals
into dEs  (e.g., \citealt{van04}). Another scenario, the so-called
harassment \citep{Moore96}, can transform  infalling late-type spirals
through close encounters with massive cluster members, and the studies
of \citet{Jerjen00}, \citet{Barazza02}, and \citet{Rijcke03} also
suggest the possibility of production of dEs by harassment.

A recent systematic study by \citet[hereafter L06 and L07,
respectively]{Lisker06a,Lisker07} 
revealed a striking heterogeneity of the class of early-type
dwarfs. They found different subclasses, with significantly different
shapes, colours, and spatial distributions. Those dEs with a disc
component have a flat shape and are predominantly found in the
outskirts of the cluster, 
suggesting that they -- or their progenitors -- might have just
recently fallen into the cluster environment. In contrast, dEs with a
compact stellar nucleus follow the classical picture of dwarf
ellipticals: they are spheroidal objects that are preferentially found
in the dense cluster center. Thus, given the structural heterogeneity
and the scatter in their stellar population characteristics, the
question arises whether they have a similar or different formation
history? Here we address this question with exploring the
relation between their stellar populations, structural parameters, and
the environment.

The paper is structured as follows. In Section 2, we
provide an overview of the data, the sample selection, and
the observational parameters. In Section 3, we present the
basic data reduction process. We describe in Section 4 the
steps followed towards the Lick index measurement, including
consistency checks of the models used, as well as tests and
comparisons of Lick indices at different spectral resolutions. The
derivation of stellar population parameters is outlined in Section 5. 
 The
results from our analysis are presented in Section 6, separating the
actual index measurements (Section 6.1) and the subsequently derived
ages and metallicties  (Section 6.2). Finally, Section 7
presents the discussion of the results, and our conclusions.

\section{Sample and Observations}

\subsection{Sample selection}
\label{sec:sub_sample}

Among the certain cluster members in the Virgo cluster catalogue
\citep[VCC,][]{Binggeli85, Binggeli93},
we selected nucleated\footnote{
The classification of a dE as nucleated
  and non-nucleated is not unambiguous. Many 
  Virgo dEs classified as non-nucleated in the VCC actually
  host a faint nucleus hardly detectable with the VCC data
\citep{Grant2005,Cote2006}.
A more appropriate term might be \emph{dEs without a nucleus of
  significant relative brightness} as compared to the central galaxy
light.}
 dEs (\,dE(N)s\,) of different morphology,
according to L06. Our target list comprises 8 dEs with discs
(\,dE(di)s\,), 7 dE(N)s without discs located in the central region of
the cluster, and 11 dE(N)s without discs located in the outer cluster
regions (including 4 objects in the dense southern subcluster). The
majority of dE(di)s populate the outer cluster regions (L07) and
therefore need to be compared to similarly located dE(N)s not hosting
a disc. One of the dE(di)s of our sample is known to host young stars
in its central region, reflected in the blue central colour \citep[VCC
0308, see][]{Lisker06b}.

One major goal of our observations was to be able to analyse the
nuclei themselves, which will be subject of a second publication to
follow. Therefore, our target selection was guided by the estimated
ratio of nucleus light to underlying galaxy light, and by the SNR of
nucleus and host galaxy. Our sample is therefore mainly biased towards
galaxies with relatively bright nuclei, both in absolute terms and
relative to the galaxy's central region. We will take this into
account in the discussion section. 
\begin{figure}
\includegraphics[width=84mm]{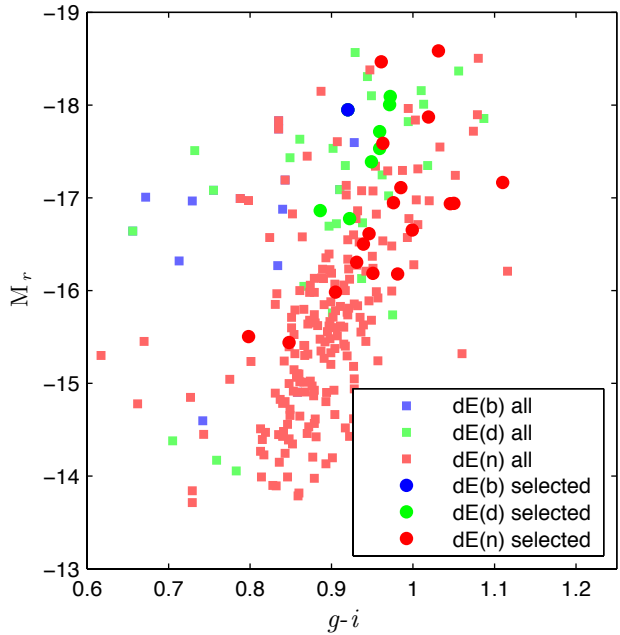}
 \caption{Colour-magnitude diagram of Virgo cluster dEs with discs
     (green), without discs (red) and blue centre (blue) as identified by L07. Solid 
     circle represent the dEs of our target sample.}
 \label{sam}
 \end{figure}

Furthermore, as shown in L07, the dE(di)s have, on average, somewhat
bluer colours than the dE(N)s. This is reflected in our sample:
Fig. \ref{sam} shows that the dE(di)s tend towards the
bluer side of the colour-magnitude relation in {\it g-i}. Whether this effect
is due to age or metallicity, and whether it is caused by the presence
of a disc itself or by environmental density, will be addressed in the
discussion section. 

We believe that this is a representative sample of bright nucleated
early-type dwarf galaxies  of the Virgo cluster with different
substructural types. Table \ref{tsam} lists the sample galaxies
and some of their basic properties.
We also build an extended sample by adding Virgo dEs from the M08
sample whose morphology is defined in L06, and remeasuring their
stellar population parameters from the reduced spectra, which were
kindly provided by C.\ Conselice on behalf of the MAGPOP ITP\footnote{http://www.astro.rug.nl/$\sim$peletier/MAGPOP\_ITP.html}
collaboration. M08 selected their galaxies to be brighter than
m$_{B} < \sim15$ mag, with high central surface
brightness, and all but one (VCC 0917) hosting a central nucleus.
In total, our extended sample thus comprises three dE(bc)s (of which
one also hosts a disc, as mentioned above), 13 dE(di)s, and 22 dE(N)s
that do not show disc features.
Note that we removed VCC 0917 from M08 sample
because it has a different subclass, as this galaxy is neither
discy, nucleated, nor blue-centered according to L06.

\begin{table*}
 \centering
 \begin{minipage}{150mm}
  \caption{The sample: basic properties.}
  \label{tsam}
\begin{tabular}{lcccccccc}
\hline
Name	&     RA	&   Dec	        &  Type	&  M$_{r}$ &
Density\footnote{Local projected galaxy number density}	& R$_{\rm eff}$	&    RV			&	Remark	\\
1&2&3&4&5&6&7&8&9\\
\hline
VCC0216	&  12:17:01.10	&  09:24:27.13	&  d,n	&  $-$16.78  &   \phantom{0}9.4	&   13.3  & 	1281 $\pm$	26	&		\\
VCC0308	&  12:18:50.90	&  07:51:43.38	&  b,d,n&  $-$17.95  &   \phantom{0}3.6	&   18.7  &	1596 $\pm$	39	&	also in M08	\\
VCC0389	&  12:20:03.29	&  14:57:41.70	&  d,n	&  $-$18.00  &   15.6	&   17.2  &	1364 $\pm$	09	&		\\
VCC0490	&  12:21:38.77	&  15:44:42.39	&  d,n	&  $-$18.09  &   31.6	&   27.6  &	1267 $\pm$	12	&		\\
VCC0545	&  12:22:19.64	&  15:44:01.20	&  n	&  $-$16.61  &   52.6	&   13.3  &	1207 $\pm$	12	&		\\
VCC0725	&  12:24:24.23	&  15:04:29.17	&  n	&  $-$16.19  &   26.2	&   25.2  &	\phantom{00}1854 $\pm$	110$^{*}$	&		\\
VCC0856	&  12:25:57.93	&  10:03:13.54	&  d,n	&  $-$17.71  &   18.2	&   15.9  &	1025 $\pm$	10	&	also in M08	\\
VCC0929	&  12:26:40.50	&  08:26:08.60	&  n	&  $-$18.58  &   26.1	&   20.5  &	\phantom{0}910 $\pm$	10	&		\\
VCC0990	&  12:27:16.94	&  16:01:27.92	&  d,n	&  $-$17.39  &   14.8	&   \phantom{0}9.9  &	1727 $\pm$	34	&	also in M08	\\
VCC1167	&  12:29:14.69	&  07:52:39.22	&  n	&  $-$16.95  &   45.1	&   27.3  &	\phantom{00}1980 $\pm$	240$^{*}$	&		\\
VCC1185	&  12:29:23.51	&  12:27:02.90	&  n	&  $-$16.65  &   64.1	&   19.3  &	\phantom{0}500 $\pm$	50	&		\\
VCC1254	&  12:30:05.01	&  08:04:24.18	&  n	&  $-$17.17  &   68.9	&   14.9  &	1278 $\pm$	18	&		\\
VCC1261	&  12:30:10.32	&  10:46:46.51	&  n	&  $-$18.47  &   21.0	&   22.5  &	1871 $\pm$	16	&	also in M08	\\
VCC1304	&  12:30:39.90	&  15:07:46.68	&  d,n	&  $-$16.86  &   15.1	&   16.2  &	\phantom{0}$-$109 $\pm$	78$^{*}$	&		\\
VCC1308	&  12:30:45.94	&  11:20:35.52	&  n	&  $-$16.50 &   30.2	&   11.4  &	1721 $\pm$	45	&		\\
VCC1333	&  12:31:01.07	&  07:43:23.04	&  n	&  $-$15.44 &   24.5	&   18.5  &	1251 $\pm$	28	&		\\
VCC1348	&  12:31:15.73	&  12:19:54.38	&  n	&  $-$16.94  &   52.4	&   13.1  &	1968 $\pm$	25	&		\\
VCC1353	&  12:31:19.45	&  12:44:16.77	&  n	&  $-$15.51  &   36.6	&   \phantom{0}8.8  &	\phantom{0}$-$384 $\pm$	82$^{*}$	&		\\
VCC1355	&  12:31:20.21	&  14:06:54.93	&  n	&  $-$17.59  &   29.9	&   29.6  &	1332 $\pm$	63	&		\\
VCC1389	&  12:31:52.01	&  12:28:54.53	&  n	&  $-$15.98  &   42.9	&   12.8  &	\phantom{0}858 $\pm$	25	&		\\
VCC1407	&  12:32:02.73	&  11:53:24.46	&  n	&  $-$16.95  &   35.7	&   11.8  &	1019 $\pm$	03	&		\\
VCC1661	&  12:36:24.79	&  10:23:05.25	&  n	&  $-$16.18  &   13.0	&   18.9  &	1457 $\pm$	34	&		\\
VCC1826	&  12:40:11.26	&  09:53:45.99	&  n	&  $-$16.30  &   13.3	&   \phantom{0}7.8  &	2033 $\pm$	38	&		\\
VCC1861	&  12:40:58.57	&  11:11:04.34	&  n	&  $-$17.78  &   23.0	&   18.4  &	\phantom{0}629 $\pm$	20	&	also in M08    	\\
VCC1945	&  12:42:54.09	&  11:26:18.09	&  n	&  $-$17.11  &   26.6	&   21.5  &	\phantom{00}1619 $\pm$	104$^{*}$	&		\\
VCC2019	&  12:45:20.44	&  13:41:34.50	&  d,n	&  $-$17.53  &   \phantom{0}7.8	&   18.1  &	1895 $\pm$	44	&		\\
\\
\multicolumn{7}{l}{{\it Galaxies from \citet{Michielsen08} incorporated in our extended sample}:}\\                       
VCC0021	&  12:10:23.15	&  10:11:19.04	&  b,n  &  $-$17.01 &   \phantom{0}1.8	&   15.2  &     \phantom{0}486 $\pm$ 25 & \\
VCC0397	&  12:20:12.18	&  06:37:23.56	&  d,n  &  $-$16.76  &    ---	&   12.4  &    2471 $\pm$ 46 & poss. mem. \\
VCC0523	&  12:22:04.14	&  12:47:14.60	&  d,n  & $-$18.57  &   20.2	&   25.7  &    1981 $\pm$ 20 & \\
VCC0917	&  12:26:32.39	&  13:34:43.54	&  -    &  $-$16.55  &   45.9	&   \phantom{0}9.2  &    1238 $\pm$ 20 & removed\\
VCC1087	&  12:28:14.90	&  11:47:23.58	&  n    &  $-$18.50  &   54.9	&   28.4  &     \phantom{0}675 $\pm$ 12 & \\
VCC1122	&  12:28:41.71	&  12:54:57.08	&  n    &  $-$17.13  &   66.0	&   19.1  &     \phantom{0}476 $\pm$ 10 & \\
VCC1183	&  12:29:22.51	&  11:26:01.73	&  d,n  &  $-$17.82  &   28.2	&   19.6  &    1335 $\pm$ 12 & \\
VCC1431	&  12:32:23.41	&  11:15:46.94	&  n    &  $-$17.72  &   30.2	&   \phantom{0}9.3  &    1505 $\pm$ 21 & \\
VCC1549	&  12:34:14.83	&  11:04:17.51	&  n    &  $-$17.24  &   28.3	&   10.9  &    1357 $\pm$ 37 & \\
VCC1695	&  12:36:54.85	&  12:31:11.93	&  d    &  $-$17.63  &   21.4	&   22.8  &    1547 $\pm$ 29 & \\
VCC1910	&  12:42:08.67	&  11:45:15.19	&  d,n  &  $-$17.86  &   28.1	&   13.6  &     \phantom{0}206 $\pm$ 26 & \\
VCC1912	&  12:42:09.07	&  12:35:47.93	&  b,n  &  $-$17.83  &   19.6	&   22.3  &    $-$169 $\pm$ 28 & \\
VCC1947	&  12:42:56.37	&  03:40:36.12	&  d,n  &  $-$17.61  &    ---	&   \phantom{0}9.4  &     \phantom{0}974 $\pm$ 19 & poss. mem. \\
 \hline
 \end{tabular}
In column 1 we give the galaxy name according to the VCC. Column 2
  and 3 represent the galaxy position\\ in RA and DEC (J2000.0), respectively. For column 4 we
take the morphological classification given by  L07 (with b=blue-centre,
d=discy n=nucleated). Column 5 gives the galaxy absolute magnitude in the SDSS
r-band taking m$-$M = 31.09  \citep{Mei07}, corrected for Galactic extinction (L07). Column 6 and 7 represent the local projected density
(number per sq.deg., calculated from a circular projected area
enclosing the 10th neighbor), and the effective radius in arcsec measured
 by L07, respectively. In column 8 we give the radial velocity in km/s provided
by NED; for those galaxies whose radial velocity is not listed in
NED$^{*}$ we measured their radial velocity from our data set, albeit
with larger errors due to the low spectral resolution. In the last
column we provide comments and indicate galaxies in common with the
study of M08.
\end{minipage}
\end{table*}  

\subsection{Observational characteristics}

The observations were carried out over six half nights from March 16 to 22, 2007,
with ESO VLT UT1/FORS2 \citep{Appenzeller98} in multi object
spectroscopy (MXU) mode. The dates and other observational information
are provided in Table \ref{obd}. In this table, the last two columns
give the number of exposures, the exposure time, and the resulting
SNR per pixel of the combined spectra, measured at 5000 \AA\ rest frame
wavelength after co-adding them. Spectrophotometric standard stars
were observed each night. The typical value of the seeing FWHM during the galaxy
exposures was 1.3'' as measured by Gaussian fitting of the field
stars. The detector in FORS2 comprises two 2k $\times$ 4k MIT CCDs, with a pixel
size of 15 $\times$ 15 $\mu$m$^{2}$. The standard resolution
collimator and the 2-pixel binned read-out yield a image scale
of 0.25''pixel$^{-1}$. The spectrograph slit width was chosen to be
1'', and covered 40'' in length, using the GRIS300V grism providing a
dispersion of 3.36  \AA{}pixel$^{-1}$. This setup yields a spectral
resolution,
as measured from the FWHM of the arc lines, of  $\sim$11 \AA{} at
$\sim$5000 \AA{}, which is below the instrumental resolution of Lick/IDS system ($\sim$8.4 \AA{} at 5000 \AA) see \citealt{Worthey97}.

\begin{table}
\caption{Observation Log}
\label{obd}
\begin{tabular}{lcrrr}
\hline
Name	& Chip &	Night	&	 Exp. (s)			&	SNR(pixel$^{-1}$)	\\
\hline
VCC0216	&	1	&	2007-03-19	&	2	$\times$	660	&	46	\\
VCC0308	&	1	&	2007-03-22	&	6	$\times$	590	&	49	\\
VCC0389	&	1	&	2007-03-22	&	2	$\times$	510	&	48	\\
VCC0490	&	1	&	2007-03-18	&	3	$\times$	420	&	36	\\
VCC0545	&	1	&	2007-03-17	&	3	$\times$	680	&	43	\\
VCC0725	&	1	&	2007-03-18	&	6	$\times$	670	&	31	\\
VCC0856	&	2	&	2007-03-19	&	1	$\times$	780	&	58	\\
VCC0929	&	1	&	2007-03-19	&	1	$\times$	640	&	63	\\
VCC0990	&	1	&	2007-03-22	&	3	$\times$	660	&	61	\\
VCC1167	&	1	&	2007-03-18	&	3	$\times$	720	&	45	\\
VCC1185	&	1	&	2007-03-20	&	2	$\times$	705	&	42	\\
VCC1254	&	2	&	2007-03-18	&	2	$\times$	600	&	56	\\
VCC1261	&	2	&	2007-03-20	&	1	$\times$	840	&	71	\\
VCC1304	&	1	&	2007-03-17	&	6	$\times$	570	&	43	\\
VCC1308	&	1	&	2007-03-20	&	4	$\times$	540	&	41	\\
VCC1333	&	1	&	2007-03-17	&	4	$\times$	570	&	38	\\
VCC1348	&	2	&	2007-03-19	&	2	$\times$	450	&	43	\\
VCC1353	&	1	&	2007-03-19	&	6	$\times$	760	&	40	\\
VCC1355	&	1	&	2007-03-21	&	5	$\times$	670	&	34	\\
VCC1389	&	1	&	2007-03-20	&	6	$\times$	670	&	40	\\
VCC1407	&	2	&	2007-03-22	&	5	$\times$	680	&	46	\\
VCC1661	&	1	&	2007-03-17	&	6	$\times$	600	&	36	\\
VCC1826	&	1	&	2007-03-20	&	3	$\times$	600	&	47	\\
VCC1861	&	1	&	2007-03-21	&	2	$\times$	660	&	47	\\
VCC1945	&	1	&	2007-03-19	&	5	$\times$	705	&	43	\\
VCC2019	&	1	&	2007-03-21	&	4	$\times$	540	&	41	\\
\hline
 \end{tabular}
\end{table}   

\section{Data reduction}
The first step in the data reduction was performed with the VLT FORS2
Pipeline {\sc
  Gasgano}\footnote{http://www.eso.org/sci/data-processing/software/gasgano}
implemented on the {\mbox{\sc
  EsoRex}}\footnote{http://www.eso.org/sci/data-processing/software/cpl/esorex.html}
data reduction package. This software performs the following basic
tasks: bias subtraction, aperture identification, flat-fielding,
cosmic-ray removal, and wavelength calibration (using the arc
  lines of the HeÐNe lamp).  The recipe \emph{fors\_calib} is
  used for the definition of the extraction mask based on the flat field
  and arc lamp exposures, and is also used for the creation of the
  normalized flat and master bias. In addition to that,
  \emph{fors\_calib} also calculates the dispersion coefficients which
  are used for wavelength calibration during the final reduction of the
  science frames using the second recipe \emph{fors\_science}. The
  typical uncertainties of the wavelength solution were around 0.2 pixels or less, which is
  a good agreement with the FORS pipeline standard.

  Since our spectra
were taken in MXU mode, which allows a maximum slit length of 40'',
the slit does not cover the full extent of our target galaxies \emph{and} a
sufficient region of the sky at the same time; instead, additional
slits were placed on blank sky regions. For this reason, the FORS2
Pipeline does not provide a good result for the sky background subtraction for this type
of spectra of spatially extended objects (as also mentioned in the
pipeline manual), and left significant residual signatures of sky
lines. Therefore, the night sky line removal has been done with
{\sc iraf}\footnote{Image Reduction \& Analysis Facility Software
  distributed by National Optical Astronomy Observatories, which are
  operated by the Association of Universities for Research in
  Astronomy, Inc., under co-operative agreement with the National
  Science Foundation} using the {\it background} task implemented on
the {\it
  oned} package. Inherently, residuals and uncertainties from the sky
subtraction could affect the line strength measurements that we will
obtain from these data, so we have investigated  thoroughly how this step is
implemented. The sky background is obtained from the blank slits,
which were placed along the extension of the galaxy slit.  All the
exposures have at least one blank slit for the sky background, in most
cases more than one. Nevertheless,
we found that the resulting sky-subtracted galaxy spectrum
was almost completely insensitive to how the sky region was chosen,
and whether one or two slits were used to sample the background. We
have therefore chosen the sky regions using the full length of the
blank slit, typically 38'', omitting 1''
at either side of slit.

The photometric standard
stars were observed in MOS mode. Here, the FORS2 pipeline sky
subtraction did not leave any residuals, since the target is a point
source and a sufficient amount of blank pixels are thus covered by the
slit. 

Once the galaxy frames
were fully reduced, one-dimensional spectra were extracted by summing
the pixel values in spatial direction along the central four
arcseconds, corresponding to 320 pc with a distance modulus
$m-M=31.09$ (16.5 Mpc, \citealt{Mei07}).
This is done in order to guarantee that the resulting Lick
indices will have sufficiently small errors to allow a precise stellar
population analysis, and to enable a direct comparison with
the sample of M08, who used the same
extraction aperture. This allows us to build an
extended sample (see Sect.~\ref{sec:sub_sample}) by combining their
and our galaxies.
 
\begin{figure*}
  \includegraphics[width=15cm]{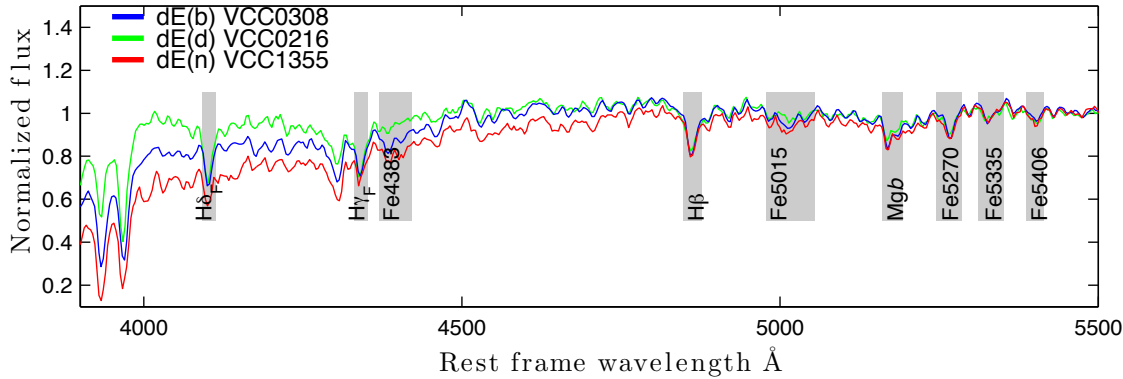}
 \caption[fspt]{Flux calibrated spectra of three typical galaxies in our sample
are shown. The specatra are shifted to restframe wavelength and 
normalized at 5500 \AA. Selected, important Lick/IDS
indices central band pass are indicated by the gray shaded region.}
 \label{fspt}
 \end{figure*}

The spectrophotometric calibration was done   
using the observations of 5 standard stars (LTT4816, LTT6248, LTT1215, LTT2415, and Feige56). The
instrument response function was computed
from the ratio of the observed standard star spectrum and the
tabulated calibrated spectrum, and the resulting continuum correction
was subsequently applied to the galaxy spectra. To
correct for the effect of atmospheric extinction, we used the {\sc iraf} task
{\it setairmass} to define the effective airmass for each
spectrum. Then, we applied the extinction correction (task {\it calibrate})
using the appropriate atmospheric extinction table (provided by Sabine Moehler). The spectral sensitivity
functions as derived from the different stars varied by typically
$\sim$1\%, reaching $\sim$3\% towards the edges of the wavelength
range. Given these small variations, we created a single function from
all standard stars, and are reasonably confident that the resulting
internal relative flux calibration is good to $\sim$1\%.

The individual spectra from different exposures of a given galaxy were then
co-added using the {\sc iraf} task scombine to produce a spectrum of
higher SNR. Three typical flux calibrated combined galaxy spectra
are shown in Fig. \ref{fspt}.

For the later calculation of index uncertainties, we also produced
error frames, following the description by
\citet{Cardiel98}, in particular the processing of error frames
parallel to the reduction steps. In this sense, we have an error
spectrum for each fully processed science spectrum that contains the
propagation of the initial random errors (due to photon statistics and
read-out noise) carried throughout the arithmetic manipulations in the
reduction procedure. 

 \section{Line Indices: measurement and calibration}

In this section we provide the details of the measurement of line
strength indices in the wavelength range 4000 \AA\ to 6000 \AA. The
main step is to calculate the line strengths particularly for Lick/IDS
indices \citep{Trager98}, using one-dimensional spectra. This will allow us to compare
the results to simple stellar population models of \citet[hereafter
BC03]{Bruzual03}, \citeauthor{Thomas03} (\citeyear{Thomas03}, here
after TMB03) and AV/MILES (Vazdekis et al. 2009; in preparation)\footnote{http://www.iac.es/galeria/vazdekis}, in order to derive SSP-equivalent ages and metallicities. In this work, we focus on maximizing the SNR of the spectra and on the
ability to combine our dataset of 26 galaxies with the 12 of
M08, who obtained their 1-D spectra by summing the
central 4 arcsec in spatial direction. We therefore apply the same extraction to
our spectra, and defer any study of potential population gradients to
a later publication.
\citeauthor{Lisker08} stated that a definite
  conclusion, as to whether the observed colour differences of different
  morphological subtypes are caused by age or metallicity effects, was
  not possible. With our good SNR in the central region, we can
  address this issue in a more appropriate and reliable way, by
  comparing our index measurements with stellar population models, in
  order to break the degeneracy of age and metallicity.

\subsection{Line strength Measurement}

The line index strengths were measured  from flux calibrated spectra using the routine {\sc Indexf}\footnote{http://www.ucm.es/info/Astrof/software/indexf/indexf.html} developed by N. Cardiel. It uses
the definition of the Lick indices from \citealt{Trager98} and calculates a
pseudo-continuum (a local continuum level) for each spectral feature
defined by the mean values within two pseudocontinuum bandpasses on either
side of the spectral feature. This software also
estimates the uncertainties resulting from the propagation of random
errors and from the effect of uncertainties on radial velocity by
performing Monte-Carlo simulations. 

Note that we have not applied a velocity dispersion correction for
our galaxies, because the expected galactic velocity dispersion,
$\sigma_{gal} \le 50$ km\,s$^{-1}$, is significantly below our spectral
resolution  $\sigma_{instr} \sim 280$ km\,s$^{-1}$. Therefore these
corrections are not necessary.

\subsection{Index Calibration and Stellar Population Model}
There are several stellar population models that provide Lick
indices for stellar populations of different ages and metallicities
(e.g, \citealt{Worthey94b}; \citealt{Vazdekis99}; BC03; TMB03). The models adopted in this
paper are those of  TMB03, BC03, and Vazdekis/MILES (new version based on MILES library,
hereafter AV/MILES). The model of TMB03 has a variable $\alpha$-element
abundance ratio, based on the standard 
SSP models computed with the code of \cite{Maraston98} with
a \citet{Salpeter55} IMF. They provide Lick index values
at Lick standard resolution. The new AV/MILES models are based
on the previous models from \citet{Vazdekis99} and \citet{Vazdekis03}. They use Padova 2000 \citep{Girardi00} isochrones. They cover
ages between 0.1 and 17.8 Gyr, and metallicities between $-$1.7 and 0.2
dex. The MILES library \citep{Sanchez06} used for these models has a
2.3 \AA{} spectral resolution (FWHM). It covers the wavelength range
from 3525 to 7500 \AA{}, with a good calibration of the flux. 

Among these three models (i.e., BC03, TMB03 \& AV/MILES), we used
  BC03 and AV/MILES for the  consistency test (see next paragraph),
  since they provide the full spectral energy distribution (SED) of
  SSPs, and can therefore be easily transformed to our spectral
  resolution of 11 \AA. Furthermore, we 
  also use AV/MILES for the correction of measured indices
  back to the Lick system (see section \ref{correction} for
  details of the correction procedure).  We use the
  $\chi^{2}-$technique to derive stellar population estimates from
  AV/MILES at 11 \AA{} resolution and TMB03 at Lick resolution (see
  section \ref{computation}). Finally, after confirming the
  consistency between the stellar parameter estimates from these two
  different models, we use those obtained with TMB03 as our final
  values, since the TMB03 model also covers different [$\alpha$/Fe]
  ratios, and thus avoids potential problems of other models with
  fixed solar abundance ratio.


Before showing the main results, we provide consistency tests
between our data and the different SSP model predictions of BC03 and AV/MILES. For this purpose, the model
indices were computed at the resolution of our data ($\sim$11 \AA).  We
then create index-index plots that contain those indices sensitive to the
same chemical species \citep{Kuntschner00}. These plots provide an
idea of how accurately the models reflect the data and, in
consequence, they help us to interpret the results with respect to a
chosen model. Moreover, it is important to check whether or not significant
differences exist in the relative spectrophotometric calibrations of
the data and the stellar libraries used by the models.  

\begin{figure*} 
 \includegraphics[width=110mm]{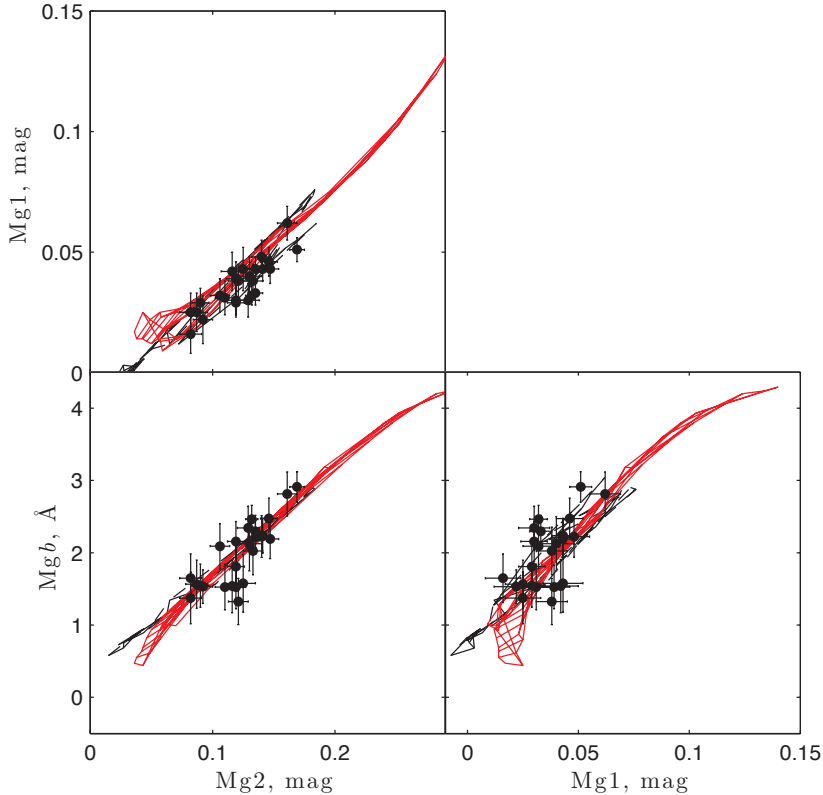}
 \caption[ctmg]{Consistency test using Mg indices and BC03 models at
   11 \AA{} resolution. The folded grids are BC03 (black) and
   Vazdekis/MILES (red) models and the datapoints correspond to the central
   indices measured in our sample. }
\label{ctmg}
 \end{figure*}

 \begin{figure*} 
 \includegraphics[width=110mm]{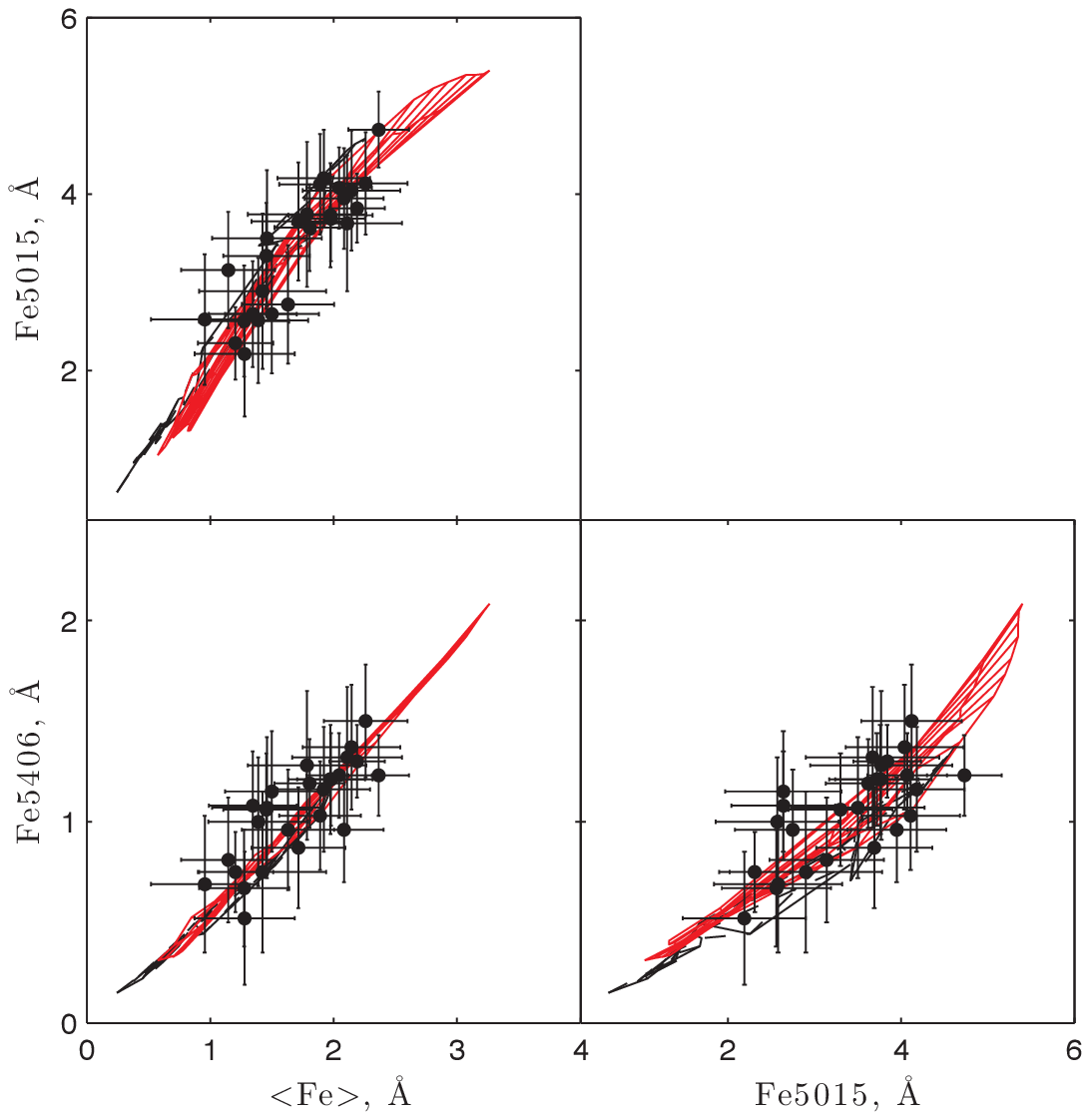}
 \caption[]{Consistency test using Fe indices and BC03 models at 11
   \AA{} resolution. The folded grids are BC03 (black) and AV/MILES (red)
   models and the datapoints correspond to the central indices measured in our
   sample.} 
\label{ctfe}
 \end{figure*}

 \begin{figure*} 
 \includegraphics[width=110mm]{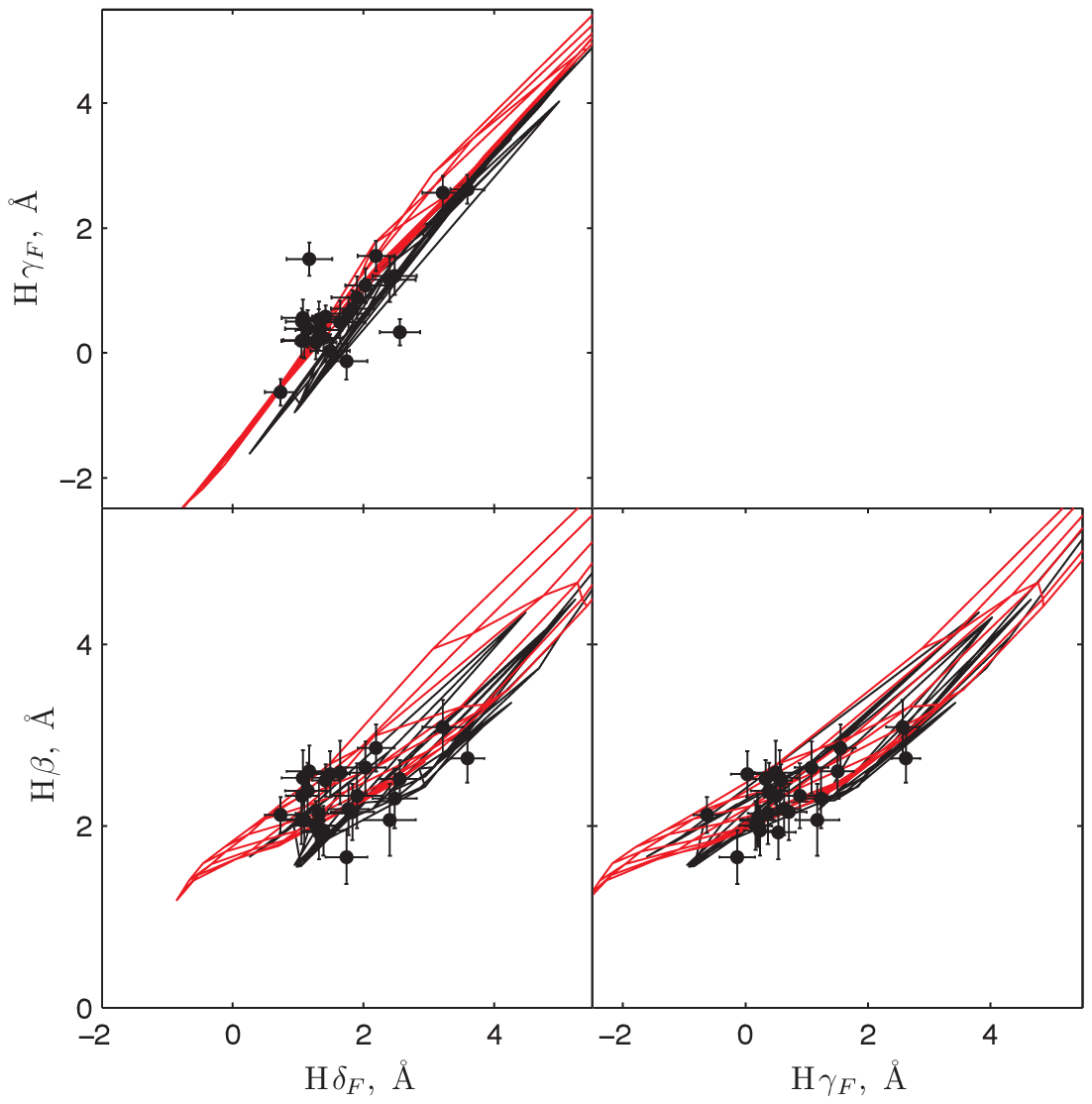}
 \caption[cthb]{Consistency test using Balmer indices indices and BC03 models at 
11 \AA{} resolution. The folded grids are BC03 (black) and AV/MILES (red)
models and the datapoints correspond to the central indices measured in our
sample.}
\label{cthb}
 \end{figure*} 
 
In Figure \ref{ctmg}, different Mg indices are plotted against each
other. The folded grids represent BC03 (black) and AV/MILES (red)
models at 11 \AA\ resolution, and the black dots are the indices measured
from our sample. The plots show a good overall agreement between the measured
index values and the model predictions, with a small systematic
deviation for galaxies with smaller Mg indices. Likewise,
Fig. \ref{ctfe} shows the relation  between $<$Fe$>$, Fe5015 and
Fe5406. In this plot, the scatter of the data points around the model
grid locus is larger; however, this is explained by our measurement
errors being larger there. Overall, the models do trace the mean trends
in the data as a whole. In Fig. \ref{cthb} we present the Balmer indices H$\beta$,
H$\delta_{F}$ and H$\gamma_{F}$. Here we find fairly good agreement
for the AV/MILES model, while a small deviation can be seen between the
BC03 model predictions and the data in the H$\delta_{F}$ and H$\gamma_{F}$
indices.

 \subsection{Correction to the Lick system}\label{correction}
\begin{figure}
 \includegraphics[width=84mm]{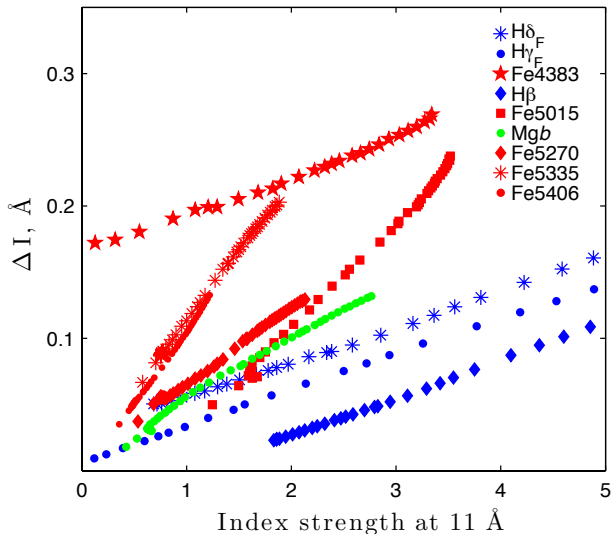}
 \caption[]{The deviation of index values after changing the
   resolution from 9\AA{} to 11\AA{} as a function of measured index
   strength at 11\AA{} of AV/MILES model spectra.}
 \label{dev}
 \end{figure}

In this section, we investigate the behaviour of line indices at different
resolutions, and provide corrections of our measured indices to the Lick
system. We assess possible effects of a variation of the resolution on the
ages and metallicities. For this
purpose, we used AV/MILES model spectra. First, we degraded the model spectra to Lick
resolution, since most studies in the literature are based on this
resolution. However, for simplicity, we consider a fixed resolution of
9 \AA, although the original
Lick system varies from $\sim$8.4 to 11.5 \AA\ depending on
wavelength. In addition, we computed another degraded version of the
model spectra, adopting the resolution of our data (11 \AA). These two
different resolutions are compared with each other in Fig.\ref{dev},
where we show the difference $\Delta$I between the measured indices for the two
resolutions (i.e., $\Delta$I = Indices measured at 9\AA\ $-$ 11\AA) as
a function of index strength. We find that Fe5335 and Fe5406 suffer
slightly more degradation in the strong absorption region than the
Balmer lines (H$\beta$, H$\gamma_F$, and H$\delta_F$) and the Mg
indices. The percentage variations of these features
are $\sim$10.5\% for Fe5335 \& Fe5406, $\sim$6\% in case of Fe5270 \&
Fe5015 and Mgb, and even smaller for H$\beta$ (i.e., $\sim$2\%) and
other higher order Balmer lines. Note that these systematic deviations
in absorption line strengths are smaller than half of the typical (statistical)
measurement error from our galaxy spectra. 

From the relations illustrated in Fig.\ref{dev}, we computed
corrections for index strength for all the
measured indices of our galaxy sample, in order to correct our data
back to the Lick resolution (9 \AA). This allows us to
apply a separate correction for each galaxy, depending on its
individual index strength. We then
further corrected each index by applying the offsets between 
flux-calibrated spectra and the non-flux-calibrated original Lick/IDS
system. For this purpose, we derived offsets from the Lick star
spectra of the MILES library (see the Section 
\ref{ap1} for more detail). With these corrections, our data qualify
for comparisons with various models based on the
Lick resolution. We are thus in the position to
compare our data to the TMB03 models, which provide tabulated values of Lick indices
as a function of age, [Z/H] and abundance ratio [$\alpha$/Fe].

\begin{figure*} 
 \includegraphics[width=170mm] {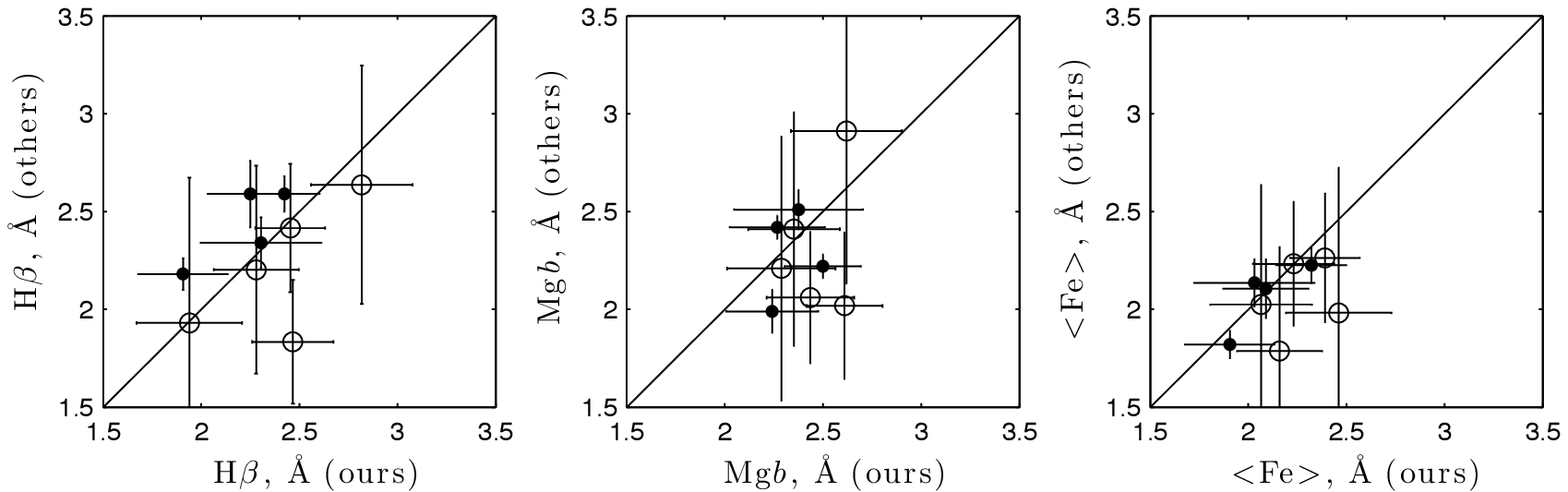}
 \caption[cpd]{Comparison of our Lick indices, corrected back to
   the Lick system, with previously published data. Open circles
   represent the galaxies in common with the M08 sample and filled
   circles stand for the data of Geha et al. (2003). }
 \label{cpd}
 \end{figure*}

To examine the robustness of our approach of correcting the
indices back to the Lick system, we compare the final corrected values
of those galaxies common to \citet{Geha03} and M08 with their published
values. In Fig. \ref{cpd}, we find that these adopted corrections give
results consistent with the previously published data.  Therefore,
our further
analyses (i.e., index analysis, comparison to models, and
extraction of age and metallicity) are done on the Lick system, unless
explicitely stated otherwise. 

 \section{Extraction of stellar population parameters from model fitting}\label{computation}

As the aim of our work is to understand the star formation
and evolutionary history of dwarf galaxies, we extract stellar
population characteristics from those indices, and combinations of
indices, that are sensitive to age, metallicity, or the relative
abundance of different metals. Population synthesis models then
provide us with SSP-equivalent, or mean
luminosity-weighted, ages and metallicities. It is necessary to mention
that the SSP-equivalent ages are biased towards younger
ages \citep{Trager00,Trager05}
since in the optical, most of the light comes from the youngest
component of the stellar population, and the age correlates most
strongly with last star formation activity. Therefore, the measured
age is influenced to a large extent by the last star formation
activity in the galaxy. However, the SSP-equivalent metallicity is an
excellent tracer of the light-weighted metallicity
\citep{Trager09}. This is because hot, young stars contribute little
to the metal lines in a composite spectrum \citep{Trager05,Serra07}.  

To convert our measured indices to ages and metallicities, we used the
$\chi^{2}-$minimization method as suggested by
\citealt{Proctor04}. In this method, we have the advantage to use a
large set of indices to get the best fitting stellar population
parameters. Therefore, it makes better use of the information that
directly depends on age, metallicity and $\alpha-$element abundance
ratio, as compared to an analysis of single indices or index pairs. To
perform the 
model fits, we first interpolated the model grid of TMB03 to a finer
grid. The actual $\chi^2-$minimization was then performed on the age,
[Z/H], and [$\alpha$/Fe] parameter space. While the method allows
us to use as many indices as possible, we used only the nine indices
with the best measurement quality
(H$\delta_F$, H$\gamma_F$, Fe4383, H$\beta$, Fe5015, Mgb,
Fe5270, Fe5335 \& Fe5406) among the measured Lick indices from our
spectra, which are tabulated in Table \ref{idm}. To calculate the 
uncertainties in the stellar population parameters, we transformed the
index error to the stellar population error contours and obtained the
maximum of the error from that.

 \begin{figure*} 
\includegraphics[width=14cm]{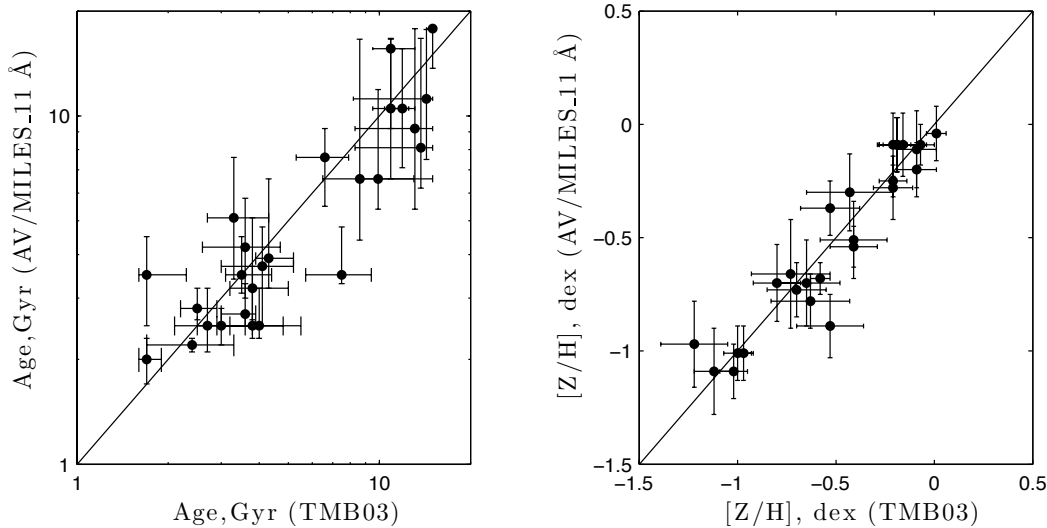}
\caption[fitid]{ Comparison between derived age \& metallicity using
  the TMB03 model at Lick resolution and the AV/MILES model at our
  resolution, i.e.,\
  11 \AA. } 
\label{comp_mes}
 \end{figure*}

In Fig. \ref{comp_mes} we show the comparison of derived ages and
metallicities using different SSP models (TMB03 \& AV/MILES). We
derived TMB03 ages and metallicities by correcting 
our data to the Lick system as described above, while AV/MILES ages
and metallicities were obtained directly at the resolution of our data
(11 \AA), by degrading the AV/MILES model down to this resolution
prior to the model fitting. We use the same set of nine indices in
both cases. Although, the measured age and metallicity agree well within the error limit, it seems that the majority of derived ages are  overestimated, when we used the model TMB0 \footnote{As a
  further test to check the source of this slight offset, we
  compared the estimated SSP parameters from the model AV/MILES at
  different resolutions (i.e., at 9\AA{} and 11\AA{}) for 
  the M08 sample. While no strong deviation is seen, there might be a small systematic bias of having younger
  ages when we degrade the resolution (the 
  plot is given in Appendix \ref{ap3}, Fig. \ref{cms_dms}). In any
  case, the estimated ages at different resolution agree well within
  the errors.}.

\begin{figure*} 
\includegraphics[width=150mm]{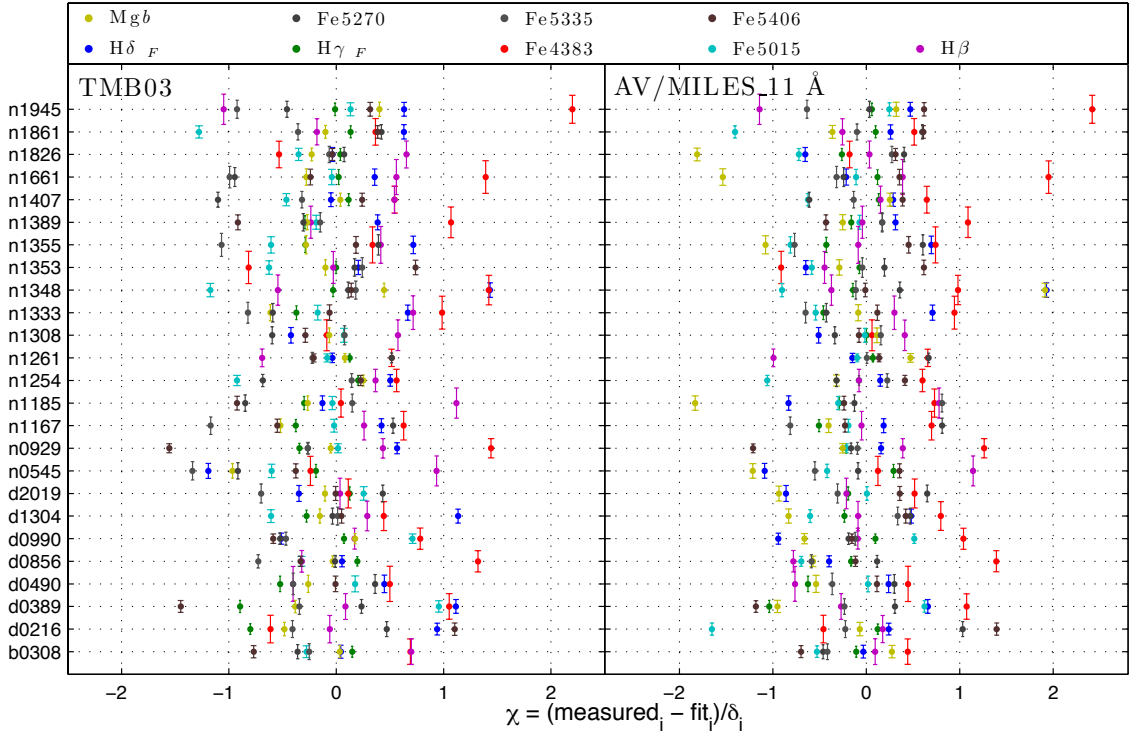}
\caption[fitid]{Distribution of the deviation of indices from the best fit model point. The Y-axis is labeled with the VCC number of our target galaxies preceded by a letter indicating their galaxy type. }
\label{fit_id}
 \end{figure*}

The behavior of the nine well-measured indices with respect to the
best-fit model index value is shown in Fig. \ref{fit_id}. Each colour
represents a different index; the galaxy number along with its subtype
label is provided as y-axis. Clearly,
the index values lie well within the 2$\sigma$ uncertainty range,
and surprisingly, in the vast majority of cases even within the 1$\sigma$ limit. It is
interesting that the deviation of the Mgb indices for the AV/MILES model is
systematically negative in almost all cases, and only for VCC 1348 it
shows a significantly positive deviation. This galaxy has the highest
$\alpha-$element abundance among our sample (see Table \ref{agm}). 
On the other hand, for the TMB03 model, the distribution of Mgb
indices is rather symmetric, and all lie within the 1$\sigma$ error
  limit. Therefore the asymmetric distribution of Mgb for the
  AV/MILES model can be interpreted as an effect of having a fixed solar
  $\alpha-$element abundance ratio. We also note that the deviations of
  Fe5015 and Fe4383 are mostly positive, while Fe5335 and Fe5406 have
  negative deviations in most cases. However, the overall mean of the
  deviations for all indices is always less than 0.5 sigma for the model TMB03.

The derived values of age and metallicity for our dEs, along with
their uncertainties, using the TMB03 model, are tabulated in Table
\ref{agmt}. We also remeasured age and metallicity for the spectral
sample of M08, using the same nine indices as in our sample. This
guarantees a proper comparison, because M08 have used a
different approach (i.e. quadratic interpolation over the nearest SSP
model grid points in the H$\beta$-[MgFe] diagram, as suggested by the
method of \citealt{Cardiel03}). Interestingly, we obtain a
smaller error on age and metallicity, as compared to their
measurement. Nevertheless, the measured ages and metallicities match
very well (see Fig. \ref{dms}).

\section{Results}
In this section, we first show the measured indices directly,
  in correlations with galaxy absolute magnitude, and in simple diagnostic
  index vs index plots, which compare them to model grids of age and
  metallicity. Finally, we analyse the stellar population
  results obtained from model fitting.
\subsection{Lick indices for Virgo dEs} 

\begin{figure}
\centering \vspace{0.0cm}
 \includegraphics[width=84mm]{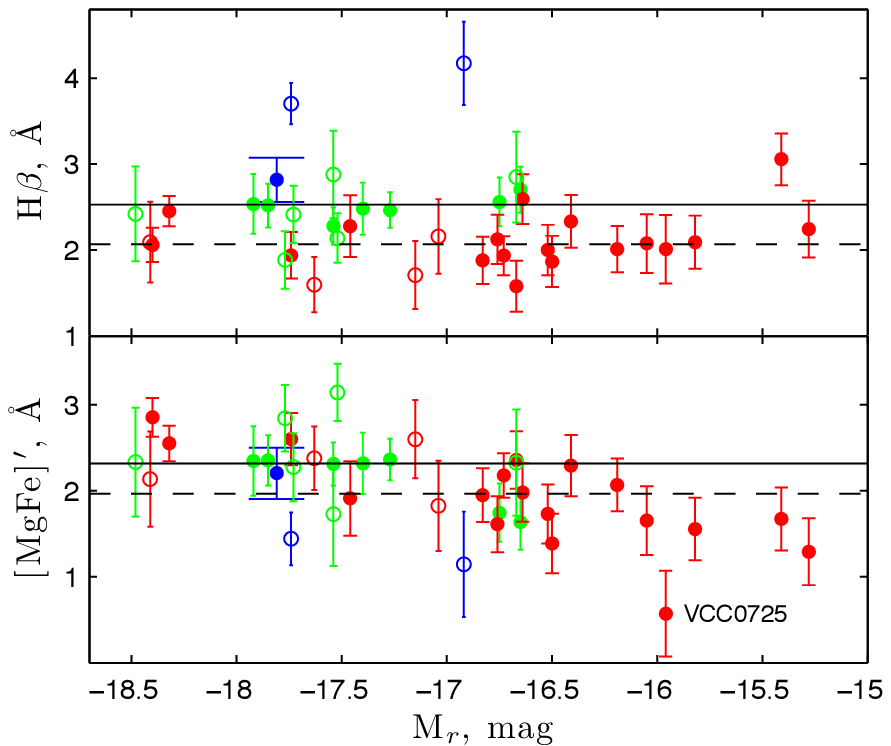}
 \caption[]{Relations between M$_{r}$ and the indices used as
   age and metallicity indicators (dE(bc): blue, dE(di): green, dE(N):
   red). Open circles represent dEs from the M08 sample. The
     median of the indices is represented by the dashed horizontal
     line for the dE(N)s and by the solid line for the dE(di)s.}
 \label{indv}
 \end{figure}

The relation of the galaxy magnitude and the indices that are widely
used as age and metallicity 
indicators (i.e.\  H$\beta$ as good age indicator and [MgFe]$^{\prime}$ as good
mean metallicity indicator, with [MgFe]$^{\prime}$ = $\sqrt{\mbox{Mgb $\times$ (0.72 $\times$ 
  Fe5270 + 0.28 $\times$ Fe5335}})$ (\citealt{Thomas03})) are shown in Figure \ref{indv}. It is
remarkable that we do not see any correlation between the age
sensitive index H$\beta$ and magnitude, while a small but clear
offset between nucleated early-type dwarfs with and without discs
(dE(di) and dE(N), respectively) is seen. On the other hand, the
metallicity sensitive index [MgFe]$^{\prime}$ shows a weak
anti-correlation with M$_{r}$, from the brighter dE(di)s down to
the fainter dE(N)s. As we
know \citep{Worthey94}, there are no such pure indices which 
only depend on either age or metallicity. Therefore, the exploration of
age and metallicity by comparing the indices or their combinations
 to model predictions is presented further below.

VCC 0725 is an outlier, due to its fairly low Mgb content (see Table A2). This galaxy is one of the
faintest among our sample and there is almost no galaxy light
beyond the central nucleus. Therefore, we suspect that a fairly large
domination of the residual sky noise within the four arcsec aperture could
produce such an effect. Hence, we
remove this galaxy from the sample during the subsequent analysis.

\begin{figure*}
 \includegraphics[width=17cm]{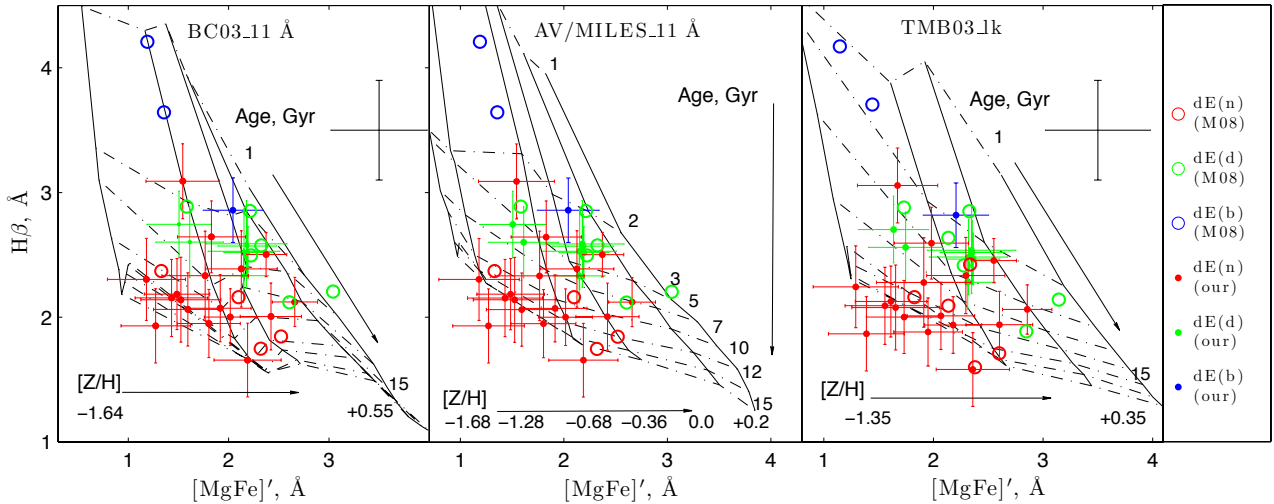}
 \caption[gridp]{The age-sensitive index H$\beta$ as a function of the
   metallicity-sensitive index [MgFe]$^{\prime}$. Overplotted are the stellar
   population models of BC03(left), AV/MILES (middle) \& TMB03 with
   [$\alpha$/Fe]=0.0 dex (right). The points without error bars are from
   the M08 sample; the mean error for these data is shown as a cross in
   the upper right corner of each panel. Colours are the  same as in
   Fig.\ref{indv}.}
\label{gridp}
 \end{figure*}

Figure \ref{gridp} shows the distribution in H$\beta$ versus [MgFe]$^{\prime}$
of dEs and a grid of simple stellar population models of BC03, AV/MILES
\& TMB03 (with [$\alpha$/Fe] = 0.0 dex).
We use [MgFe]$^{\prime}$ as
a metallicity indicator, which is considered independent of
[$\alpha$/Fe]-abundance, and H$\beta$ as age indicator, because its age
sensitivity is greater and it is less degenerate with metallicity or
abundance ratio variations than the other Balmer indices
\citep{Korn05}.

The solid lines, which are close to vertical, are of constant metallicity, whereas
  the dashed lines, which are close to horizontal, connect
constant-age models (from top to bottom: 1, 2, 3, 5, 7, 10, 12, \& 15
Gyr). All three panels appear to be fairly consistent with each
other. However, there are some galaxies that lie outside of the model
grid, located at low H$\beta$, i.e., at the bottom left corner of the
grid. Nevertheless, when taking into account their errorbars, they
could still lie within the grid region. Moreover, it is not surprising
that some galaxies (particularly dEs) fall outside of model grid. The
same problem has already been  noticed  in the similar type of study of
the dEs in the Coma cluster by \citet{Poggianti01}. This can be
an issue of how the mass loss along the red giant branch is treated in
the models, as discussed by \citealt{Maraston03}. 

We can see that most of the dEs appear to have sub-solar metallicity,
supporting the results of \citet{van04}, M08, \citet{Geha03}, \citet{Koleva09} . On the other hand,
their ages span a wide range. It is apparent that all dEs with blue
central colours (dE(bc), blue symbols) are located towards fairly young ages,
falling above the 3 Gyr line in the figure. There is
almost no overlap between the dE(bc) and the dE(N) (red symbols),
except for VCC 1353. Furthermore, most
of the discy dEs (dE(di), green in colour) lie in between dE(N) and
dE(bc). Thus, a clear division of the different subclasses in
terms of age is present, with low, intermediate and high
ages for dE(bc), dE(di) and dE(N), respectively. Despite of having a
much larger spread, the dE(N)s are clustered in the bottom
left corner of the grid -- or even below the grid in some cases --
centered on the low metallicity region. Interestingly, only three
dE(di)s fall left of the metallicity line of [Z/H] = $-0.36$ dex.

\begin{figure} 
\includegraphics[width=84mm]{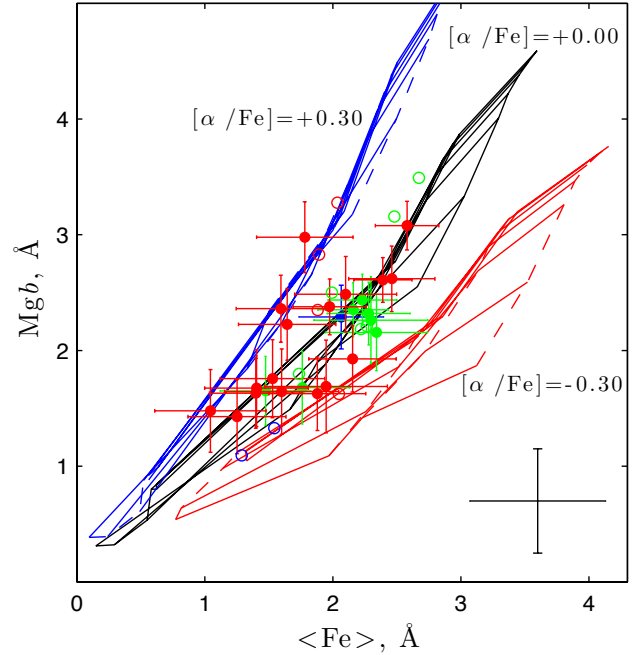}
\caption[mgfe]{Central Mg versus $<$Fe$>$ indices after correcting
  back to Lick resolution, superimposed on the TMB03 model with different
  ages, metallicities, and abundance ratios. The symbols are the same as in
  Fig.\ \ref{gridp}.}
\label{alfinm}
 \end{figure}

Fig.  \ref{alfinm} presents the Mgb vs $<$Fe$>$  diagram, where the
data are superimposed on the TMB03 model  with various
$\alpha-$element abundance ratios: [$\alpha$/Fe]= $-0.3$, 0.0, and +0.3 dex. The
results from previous studies of the stellar populations of low-mass
systems \citep{Gorgas97, Geha03, Thomas03}, that
the  $\alpha-$element abundance of dEs is consistent with solar, seem
to be in agreement with our study.
However, the scatter is fairly large for
the dE(N)s, and on average they might be slightly more $\alpha-$enhanced
than the dE(di)s.

\subsection{Age, metallicity and abundance ratios}

\begin{table}
 \caption{ Derived stellar population parameters using the model TMB03}
 \label{agmt}
\begin{tabular}{lcrr}
\hline \smallskip
Galaxy	&	Age, Gyr				&	[Z/H]	, dex		&	[$\alpha$/Fe], dex		\\
\hline
VCC0308 &	01.7	$^{	+0.2	}_{	-0.1	}$&	$-$0.09	$\pm$	0.10	&	0.01	$\pm$	0.07	\\
\hline
VCC0216	&	01.7	$^{	+0.6	}_{	-0.1	}$&	$-$0.53	$\pm$	0.17	&	0.19	$\pm$	0.13	\\
VCC0389	&	04.0	$^{	+0.8	}_{	-0.8	}$&	$-$0.21	$\pm$	0.07	&	$-$0.01	$\pm$	0.08	\\
VCC0490	&	02.7	$^{	+0.9	}_{	-0.6	}$&	$-$0.16	$\pm$	0.12	&	$-$0.04	$\pm$	0.11	\\
VCC0856	&	03.5	$^{	+0.9	}_{	-0.4	}$&	$-$0.21	$\pm$	0.07	&	$-$0.01	$\pm$	0.08	\\
VCC0990	&	03.6	$^{	+0.3	}_{	-0.6	}$&	$-$0.19	$\pm$	0.07	&	$-$0.06	$\pm$	0.08	\\
VCC1304	&	04.1	$^{	+1.1	}_{	-1.1	}$&	$-$0.65	$\pm$	0.17	&	$-$0.04	$\pm$	0.18	\\
VCC2019	&	03.0	$^{	+0.8	}_{	-0.5	}$&	$-$0.19	$\pm$	0.10	&	$-$0.12	$\pm$	0.11	\\
\hline
VCC0545	&	11.9	$^{	+0.6	}_{	-1.5	}$&	$-$0.97	$\pm$	0.05	&	$-$0.06	$\pm$	0.22	\\
VCC0929	&	04.3	$^{	+0.9	}_{	-0.4	}$&	0.01	$\pm$	0.05	&	0.00	$\pm$	0.05	\\
VCC1167	&	10.9	$^{	+2.2	}_{	-1.4	}$&	$-$1.00	$\pm$	0.07	&	0.07	$\pm$	0.16	\\
VCC1185	&	09.9	$^{	+3.1	}_{	-1.3	}$&	$-$0.80	$\pm$	0.12	&	$-$0.25	$\pm$	0.16	\\
VCC1254	&	06.6	$^{	+1.3	}_{	-1.3	}$&	$-$0.41	$\pm$	0.12	&	0.07	$\pm$	0.08	\\
VCC1261	&	02.5	$^{	+0.4	}_{	-0.3	}$&	$-$0.07	$\pm$	0.07	&	0.01	$\pm$	0.05	\\
VCC1308	&	03.8	$^{	+1.2	}_{	-0.6	}$&	$-$0.21	$\pm$	0.10	&	0.07	$\pm$	0.09	\\
VCC1333	&	13.7	$^{	+1.3	}_{	-5.4	}$&	$-$1.12	$\pm$	0.10	&	0.12	$\pm$	0.23	\\
VCC1348	&	15.0	$^{	+0.0	}_{	-0.7	}$&	$-$0.58	$\pm$	0.05	&	0.38	$\pm$	0.09	\\
VCC1353	&	02.4	$^{	+0.9	}_{	-0.7	}$&	$-$0.63	$\pm$	0.20	&	0.19	$\pm$	0.22	\\
VCC1355	&	03.8	$^{	+1.7	}_{	-0.9	}$&	$-$0.43	$\pm$	0.22	&	$-$0.16	$\pm$	0.16	\\
VCC1389	&	10.9	$^{	+2.2	}_{	-1.4	}$&	$-$1.02	$\pm$	0.07	&	0.08	$\pm$	0.18	\\
VCC1407	&	14.3	$^{	+0.7	}_{	-6.1	}$&	$-$0.70	$\pm$	0.15	&	0.13	$\pm$	0.13	\\
VCC1661	&	08.6	$^{	+6.4	}_{	-2.1	}$&	$-$0.73	$\pm$	0.20	&	$-$0.23	$\pm$	0.16	\\
VCC1826	&	07.5	$^{	+1.9	}_{	-1.8	}$&	$-$0.53	$\pm$	0.15	&	$-$0.15	$\pm$	0.14	\\
VCC1861	&	03.3	$^{	+1.0	}_{	-0.6	}$&	$-$0.09	$\pm$	0.10	&	$-$0.03	$\pm$	0.08	\\
VCC1945	&	03.6	$^{	+1.1	}_{	-1	}$&	$-$0.41	$\pm$	0.17	&	0.07	$\pm$	0.12	\\
\hline
\\
\multicolumn{4}{l}{{\it Galaxies from \citet{Michielsen08}}:}\\ 
\hline
VCC0021	&	01.0	$^{	+0.8	}_{	-0.0	}$&	$-$0.90	$\pm$	0.25	&	0.00	$\pm$	0.30	\\
VCC1912	&	02.2	$^{	+0.6	}_{	-0.7	}$&	$-$0.75	$\pm$	0.15	&	$-$0.11	$\pm$	0.19	\\
\hline
VCC0397	&	01.7	$^{	+0.6	}_{	-0.4	}$&	$-$0.12	$\pm$	0.25	&	$-$0.04	$\pm$	0.16	\\
VCC0523	&	04.1	$^{	+2.7	}_{	-1.4	}$&	$-$0.29	$\pm$	0.25	&	0.15	$\pm$	0.18	\\
VCC1183	&	03.6	$^{	+0.9	}_{	-0.9	}$&	$-$0.26	$\pm$	0.12	&	$-$0.06	$\pm$	0.11	\\
VCC1695	&	02.1	$^{	+1.5	}_{	-0.5	}$&	$-$0.38	$\pm$	0.30	&	0.04	$\pm$	0.26	\\
VCC1910	&	05.7	$^{	+2.5	}_{	-1.4	}$&	$-$0.09	$\pm$	0.10	&	0.05	$\pm$	0.08	\\
VCC1947	&	04.5	$^{	+1.2	}_{	-0.8	}$&	0.13	$\pm$	0.10	&	0.09	$\pm$	0.07	\\
\hline
VCC1087	&	03.6	$^{	+1.6	}_{	-1.2	}$&	$-$0.24	$\pm$	0.17	&	$-$0.12	$\pm$	0.16	\\
VCC1122	&	05.2	$^{	+1.7	}_{	-1.9	}$&	$-$0.61	$\pm$	0.25	&	$-$0.29	$\pm$	0.20	\\
VCC1431	&	13.1	$^{	+1.9	}_{	-2.7	}$&	$-$0.53	$\pm$	0.12	&	0.28	$\pm$	0.12	\\
VCC1549	&	04.3	$^{	+1.4	}_{	-1.5	}$&	$-$0.12	$\pm$	0.15	&	0.19	$\pm$	0.11	\\

\hline
\end{tabular}
\end{table}

Using the combined sample of our galaxies and the re-measurement of
the M08 galaxies' spectra with the method of the $\chi^{2}-$minimization, we find a mean age of
3.0 $\pm0.8$ Gyr  and a mean metallicity of
[Z/H] = $-$0.31 $\pm$0.10 dex for the dE(di)s. In contrast,
the mean values for the dE(N)s are 7.5 $\pm1.9$ Gyr and
[Z/H] = $-$0.54 $\pm$0.14 dex. The dE(bc)s have comparatively young 
ages (i.e less than 3 Gyr). Two of them show a rather low metallicity ($-0.90$ $\pm$0.25 dex for VCC0021 and $-0.75$ $\pm$0.15 dex for VCC1912), while the third one
  also belongs to the discy subclass (VCC 0308), and indeed has a
  significantly larger metallicity (i.e $-0.09$ $\pm$0.10 dex), consistent with the other
  dE(di)s. 

As a statistical comparison, we have used the Kolmogorov-Smirnov
test for the goodness of fit with the null hypothesis that two
observed distributions are from the same continuous distribution.
The estimated probability P$_{KS}$ that the difference is at least as
large as observed if they had been drawn from the same population, is
P$_{KS}$ = 0.01, 0.03 and 0.15 for the age, metallicity, and
[$\alpha$/Fe]-abundance distributions, respectively, of dE(N)s and
dE(di)s. This apparently confirms that the age and metallicity differ
significantly for dEs with and without discs. However, this
interpretation is challenged when the stellar
population characteristics are considered with respect to galaxy
luminosity; details are given at the end of this section.

\begin{figure}
\includegraphics[width=84mm]{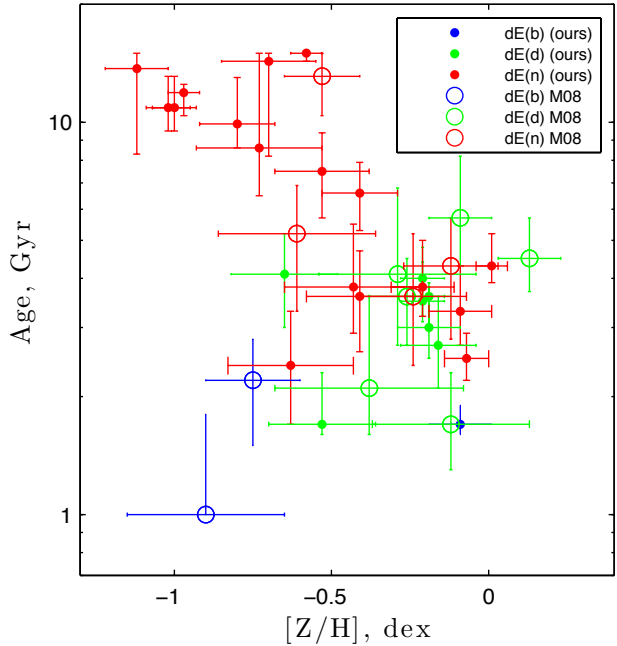}
\caption[fitid]{Age versus metallicity [Z/H]. Here colour
  and symbols are the same as in Fig.\ref{indv}.} 
\label{agm}
 \end{figure}

The galaxies in our sample follow a general trend linking age and
metallicity: as displayed in Figure \ref{agm}, we see that the derived
ages are decreasing with increasing metallicity of the dEs, similar
to the observation of \citealt{Rakos01} for Fornax dEs. 
  Interestingly, we see that an age-metallicity anti-correlation is
  present for the dE(N)s, even tighter than previously
  noticed by \citet{Poggianti01} for Coma cluster dEs.
 Such an age-metallicity anti-correlation is also found for giant
 early-type galaxies, but it is under debate because of the
 correlation of the metallicity and age errors
 \citep{Trager00,Kuntschner01}.
  Nevertheless,
  together with the dE(di)s, it can be noticed that the brighter dEs
  might even follow an opposite trend of increasing metallicity with
  increasing age. It needs to be mentioned, however, that the scatter
  of values is fairly large.

\begin{figure}
\includegraphics[width=6.5cm]{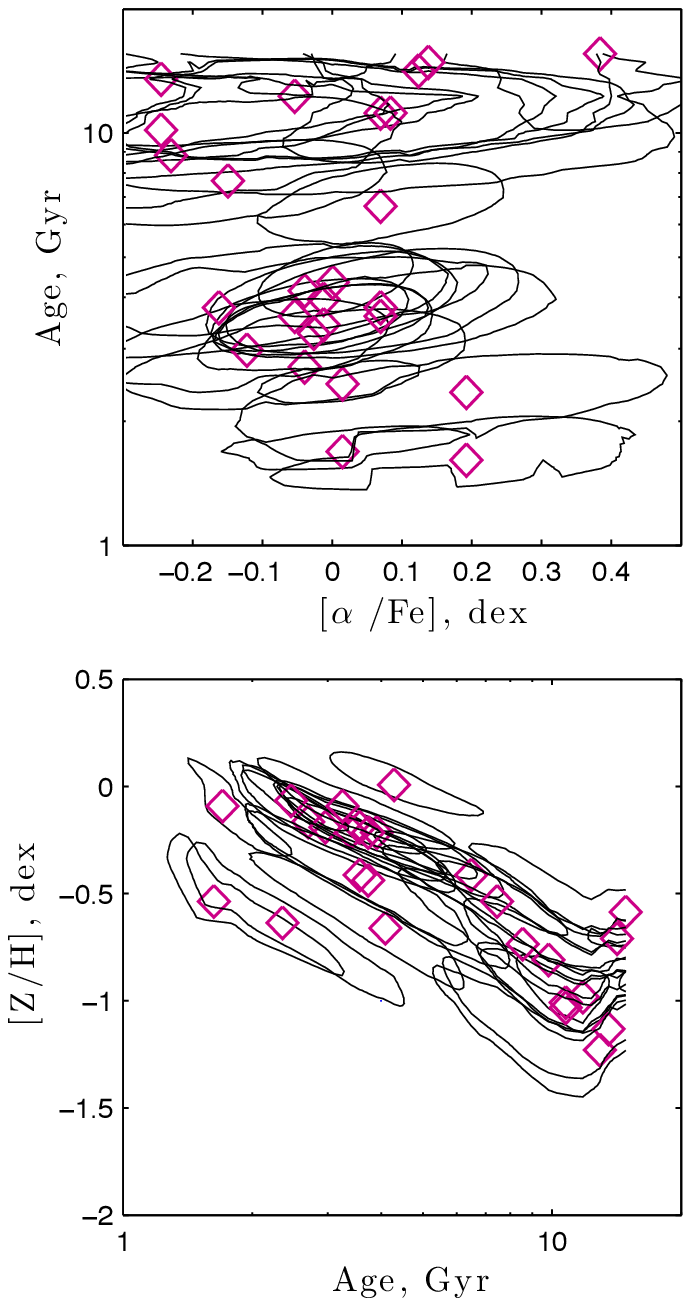}
\caption[fitid]{Error contours in different projection planes of age,
  metallicity and [$\alpha$/Fe]-abundance space.} 
\label{coun}
 \end{figure}

In addition, we provide a plot with the error ellipses
(Fig.\ref{coun}), to better illustrate the behaviour of our measurements with
respect to the age and metallicity
anti-correlation. \citealt{Kuntschner01} suggest that, when the
  measurement errors in the line strength are not negligible, the tilt
  of the model grid leads to the correlated errors in age and
  metallicity, as visualized by  the error contours in the age metallicity
  plane in Fig.\ref{coun}. 

\begin{figure*}
\includegraphics[width=15cm]{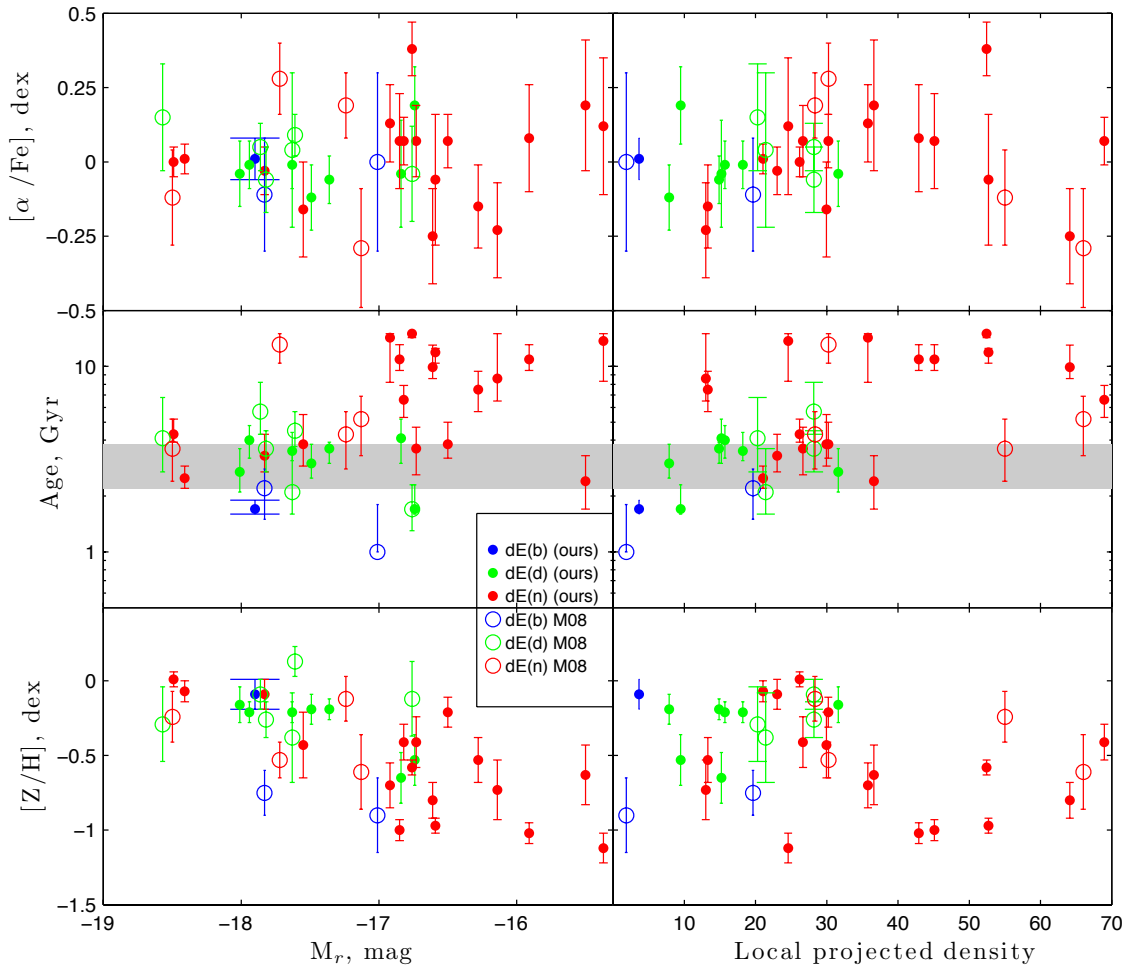}
\caption[fitid]{The derived ages (top), metallicities (middle) and
  [$\alpha$/Fe]-abundance (bottom), plotted against $r$ band absolute magnitude (left),
  and against local projected density (right, number per
  sq.deg.). Colours and symbols are the same as in
  Fig.\ref{indv}. The gray shaded region represents the mean 
$\pm\sigma$ age and metallicity of dEs(di)}
\label{env}
 \end{figure*}

The luminosity-weighted metallicities derived from the well measured
nine indices are shown as a function of galaxy absolute magnitude in Figure
\ref{env} (bottom left). Both dE(di)s and dE(N)s follow the
well known metallicity$-$luminosity relation
\citep{Poggianti01}. In contrast, we do not see any strong
correlation between measured metallicity and local projected galaxy
density (bottom right).
In the top panel of the
Figure, one can see that the dEs in general are consistent
with solar [$\alpha$/Fe]-abundance, and we do not see any relation
with luminosity or local density.

Unlike the correlation of
metallicity and luminosity, the measured ages show  an aniti-correlation
with the luminosity, and correlation with local projected density, although the latter
trend is not as strong as the former.
It appears that this might be the cause of the apparent age
difference between dEs with and without discs, since in addition to
the relation with magnitude, the dE(N)s are situated in
the denser cluster regions as compared to the dE(di)s. Thus, taking
the left and right panels in Fig.~\ref{env} together, no significant
difference is seen between dE(di)s and dE(N)s in either age or
metallicity, \emph{as long as they are compared at the same magnitude
  and local density}.

On the other hand, the bimodality in the age distribution is quite
  prominent. Interestingly, the majority of those dEs belonging to the
  group with lower ages lie consistently within the one-sigma area of
  the dE(di)s (gray-shaded region in Fig.~15). But we do not see
  any such strong bimodality in the metallicity distribution.
 
 \section{Discussion And Conclusion}
  We have undertaken a spectroscopic investigation of Virgo cluster
  dEs, deriving ages, metallicities and $\alpha-$element abundance
  ratios for 38 dEs of different types, i.e., dEs with and without
  discs, and examined the variation of their stellar populations
  according   to their morphology. We have established that the
  mean ages and metallicities of different types of dEs are different.
 
 This result mostly represents the central part of the galaxy, as we used a
 central fixed aperture to extract the one dimensional spectra of the
 galaxies, in order to optimize the SNR for our Lick index
 analysis. We expect that having a wider aperture effectively
 minimizes the influence of central nucleus. In a worst-case scenario,
 for those dEs that have a relatively low surface brightness, i.e., a
 rather faint stellar envelope compared to the nucleus, it is not
 unquestionable whether our extracted spectra do really represent the
 galaxies as a whole, or the nuclei only. Interestingly, though, we
 found that most of the faint nucleated dEs have a fairly old
 stellar population, and that typically, the nuclei tend to be younger
 than their host galaxies \citep{Paudel09}. This appears
 consistent with the studies of 
 \citet{Chilingarian09} for Virgo dEs and \citet{Koleva09} for Fornax
 cluster dEs: both concluded that the central part of the
 galaxies are relatively younger and more metal-rich than the outer part. In
 that respect, we can argue that our conclusions are not biased by
 having a strong effect of the central nucleus in the spectral
 extraction process. If anything, our result of old galaxy ages
   at fainter magnitudes could only become somewhat more pronounced by
   removing the nucleus contribution.

It was known before that the different dE subclasses, in particular
those with and without discs, exhibit differences in their colours,
magnitude, and the local environment in which they reside
\citep{Lisker07}. The dE(di)s were found to have, on average, slightly
bluer colours than the dE(N)s, and \citeauthor{Lisker07} interpreted
this colour difference with either older age or higher metallicity of
the dE(N)s. By combining our sample with that of M08, and therefore
reaching statistically significant subsample sizes, we see that the
colour offset is mainly an effect of age rather than
metallicity. However, it is not straightforward to tell which
governing factor is most responsible for this apparent correlation of
galaxy substructure and stellar population characteristics. As noted
above, we find correlations of age with both magnitude and local
density, and we know that the dE(di)s are relatively bright and 
typically located in less dense cluster regions than the dE(N)s; the
combination thus shows that no significant difference between the
stellar population of dE(di)s and dE(N)s can be claimed. This is
interesting, since \citet{Lisker08} pointed out that the fairly wide
(estimated) distribution of intrinsic axial ratios of the dE(N)s would
be consistent with assuming that the dE(di)s -- most of which are
nucleated -- are simply the flat tail of this distribution. The
location in regions of comparatively lower density could then be
interpreted such that discs in dwarfs can only survive for a
reasonably long time if the amount of dynamical heating due
to tidal forces is low \citep[also see][]{Mastropietro05}, and
therefore, dEs with discs are preferentially found outside of the
cluster center. Moreover, L06 noted that there could still be a
significant fraction of objects that have discs, but that were simply
not found by their analysis of SDSS images, due to low SNR or the lack
of disc features like spiral arms or bars. We see in our data that a
handful of dE(N)s are as young and metal rich as the dE(di)s, and
interestingly, they also lie in a low density region. So maybe, just
as a possibility, these could be those dE(N)s that actually host
discs, but were not identified as such --- which would then also
explain the bimodality when compared to the fainter ones.

Furthermore, an apparent inconsistency is observed in the relation
between age and luminosity: previous studies provided evidence for a
continuously later formation epoch and/or a longer duration of star
formation when going from the massive ellipticals to the dwarfs
\citep[see, e.g.,][]{Gavazzi02,Boselli05,LiskerHan08}, which
apparently is at variance with our findings. On the other hand, this
trend is unlikely to continue to even fainter magnitudes, since most
of the Local Group dwarf spheroidal (dSph) galaxies are know to be
dominated by old populations \citep[e.g.][]{Grebel99}.
It seems that, the fainter dEs are significantly older  than the
brighter ones, no matter at which density they are
located. To statistically confirm this result, we divide the total
sample in two groups: those being fainter and brighter than the mean
magnitude M$_{r}$ = $-$17.22 mag. We obtain a mean age of  3.9 Gyr and 7.4 Gyr for the
brighter and fainter group, respectively, which is very similar to the
mean ages of the discy and nucleated subsamples (3.0 and 7.5
Gyr). Furthermore, the K-S test supports our finding that a dichotomy
in the distribution of ages can be seen not only with respect to
different morphology but also different luminosity, as we get a K-S
test probability of 0.01 for the null hypothesis that the age
distributions of the fainter and brighter group are not different.

The ages of 10 Gyr and above indeed appear more comparable to
the Local Group dSphs.  In that
sense, we find a discontinuity in the trend of stellar population
characteristics with luminosity, and we can speculate that the fainter
dEs might be a different species than the brighter ones or simply
  were not able to sustain star formation for as long a period as the
  more luminous and presumably more massive dEs.

In comparison to previous stellar population studies of dEs in different
clusters, unlike  \cite{Chilingarian08} who found a
correlation of [$\alpha$/Fe] with the projected distance in the
Abell496 cluster -- although only in the very central part -- and the
study of \cite{Smith09} for Coma, we do not find any
correlation between $\alpha$-element abundance ratio with
environmental density. This could also be due to differences between the clusters,
in the sense that Coma is a more virialized cluster of higher richness
class than Virgo. Nevertheless, our range of 
age and metallicity roughly agrees with their ranges. Another study by
\citet{Geha03}, focusing on the stellar
populations of rotating and non-rotating Virgo dEs,
resulted in no difference between them. On the other hand, a recent,
more extensive study by \citet{Toloba2009} finds that rotating Virgo
dEs are typically located further from the cluster center and have
younger stellar population ages.
 We have only six dEs in
common with these studies (counting only those for which tabulated
values were published). Three of them (VCC 0308, VCC 0856 \& VCC 1947)
have disc substructure and
are rotating, while for three (VCC 1254, VCC 1261 \& VCC 1308) no disc
features were identified and they are non-rotating according to
\citeauthor{Geha03}. Note, though, that VCC 1261 is clearly rotating
from the analysis of dE globular clusters of \citet{Beasley09}. 
\cite{Chilingarian09} shows that this galaxy contains a
  rotating kinematically decoupled core, indicating the difficulties
  in drawing robust conclusions on the global rotation of a dE from
  the inner stellar kinematics. 
We also find that the overall stellar populations are quite mixed
within these six galaxies, as has already been noted by \citet{van04}
in their study of (non-)rotating dEs.

There have been numerous discussions on the origin of the passive
dwarf galaxies. The key issue is whether cluster dwarfs
are {\it primordial}, having formed from the highest density peaks in
the proto-cluster \citep{Tully02} and avoiding subsequent merging, or
originating from transformation of late-type disc galaxies through the
interaction with the cluster environment
\citep[e.g.][]{Boselli06}. However, to explain the formation of dEs,
using such single scenarios may be too simplistic, and it is likely
that both scenarios operate at some level. Furthermore, the question
is how to find unique features for such a process that dominates
in a given mass regime, a given environment, and for a given
morphological subclass of dwarfs.

Our results promote the idea of different scenarios for different
morphology and/or different luminosity of dEs, although the physical
mechanisms responsible for the quenching of the star formation activity
in the galaxies are not easy to understand from the stellar population
properties. The observed systematic difference in age \& metallicity
between different types of dEs support the finding of L07. The typical
age range of the brighter dEs of less than 5 Gyr indicates that the
star formation activity ceased at z $<$ 0.5. The younger age, higher
metallicity and consistency with solar $\alpha-$element abundance
would suggest that a discy dE can indeed form through the structural
transformation of a late-type spiral into a spheroidal system,
triggered by the popular scenario of strong tidal interactions with
massive cluster galaxies. A stellar disc component (which has rather high
metallicity) may survive and form a bar and spiral features that can
be retained for some time depending on the tidal heating of the galaxy 
(\citealt{Mastropietro05}; also see \citealt{LiskerFuchs2009} ). This ``galaxy harassment'' process
could thus produce disc-shaped dEs.

On the other hand, the fainter dE(N)s, with their older ages, might be a
different class of objects with a different formation
scenario. The poorer metal content supports the idea 
that they might be primordial objects, as suggested by
\citet{Rakos04}; they might have either suffered early infall into the
cluster potential, or formed together with cluster itself.
The generally low surface brightness of dEs suggests that whichever
mechanism is responsible for the halting the star formation
activity, it must have been rather efficient.
The common idea is that internal feedback might be
responsible for their cessation of star formation activity at such
early epochs. We can speculate that the slightly higher mean
[$\alpha$/Fe] as compared to the dE(di)s could indicate a more
efficient star formation activity than their discy counterparts before
the quenching. High-effiency star formation has also been
  suggested as an explanation for the comparatively high metallicity
  of the old populations of Local group dSph as compared to dIrr
  populations of the same age, i.e., as an explanation for the offset
  along the
  metallicity$-$luminosity relation \citep{Grebel03}. However, further
dynamical studies of these fainter 
galaxies are needed to prove whether the removal of gas by
internal feedback processes is efficient or not, since it has been
pointed out that it is 
difficult to really eject the gas from a dwarf galaxy by supernovae
explosions unless the dwarfs have masses less than 10$^{7}$ solar masses
\citep{MacLow99}.

\section*{Acknowledgments}
We thank the referee  Igor Chilingarian
  for providing the useful comments and suggestions that helped improving the manuscript. 
     We are indebted
to C. Conselice and the MAGPOP ITP collaboration for providing the
spectra of the M08 sample. S.P. would like to thank Anna
Pasquali \& Alexander Hansson for their helpful advice during the data
reduction and analysis.   T.L. and E.K.G. would also like to thank
Bruno Binggeli for valuable input to the observing proposal. T.L.\
and K.G.\ are grateful to Emanuela Pompei for excellent support during
the observations, and to the Swiss National Science Foundation for
financial support under grant 200020-113697. 
 We are grateful to N. Cardiel  for providing his program {\sc Indexf}
 and supporting us with its application. S.P.\ and T.L.\ are supported within the framework of the Excellence
    Initiative by the German Research Foundation (DFG) through the Heidelberg
    Graduate School of Fundamental Physics (grant number GSC 129/1).
 S.P. acknowledges 
the support of the
International Max Planck Research School (IMPRS) for Astronomy and
Cosmic Physics at the University of Heidelberg.
 This work has made
use of the NASA Astrophysics Data System and the NASA/IPAC
Extragalactic Data base (NED) which is operated by the Jet Propulsion
Laboratory, California Institute of Technology, under contract with
the National Aeronautics and Space Administration.


\bibliographystyle{mn2e}


\appendix

\section{Lick offsets} \label{ap1}

\begin{figure*}
 \includegraphics[width=175mm]{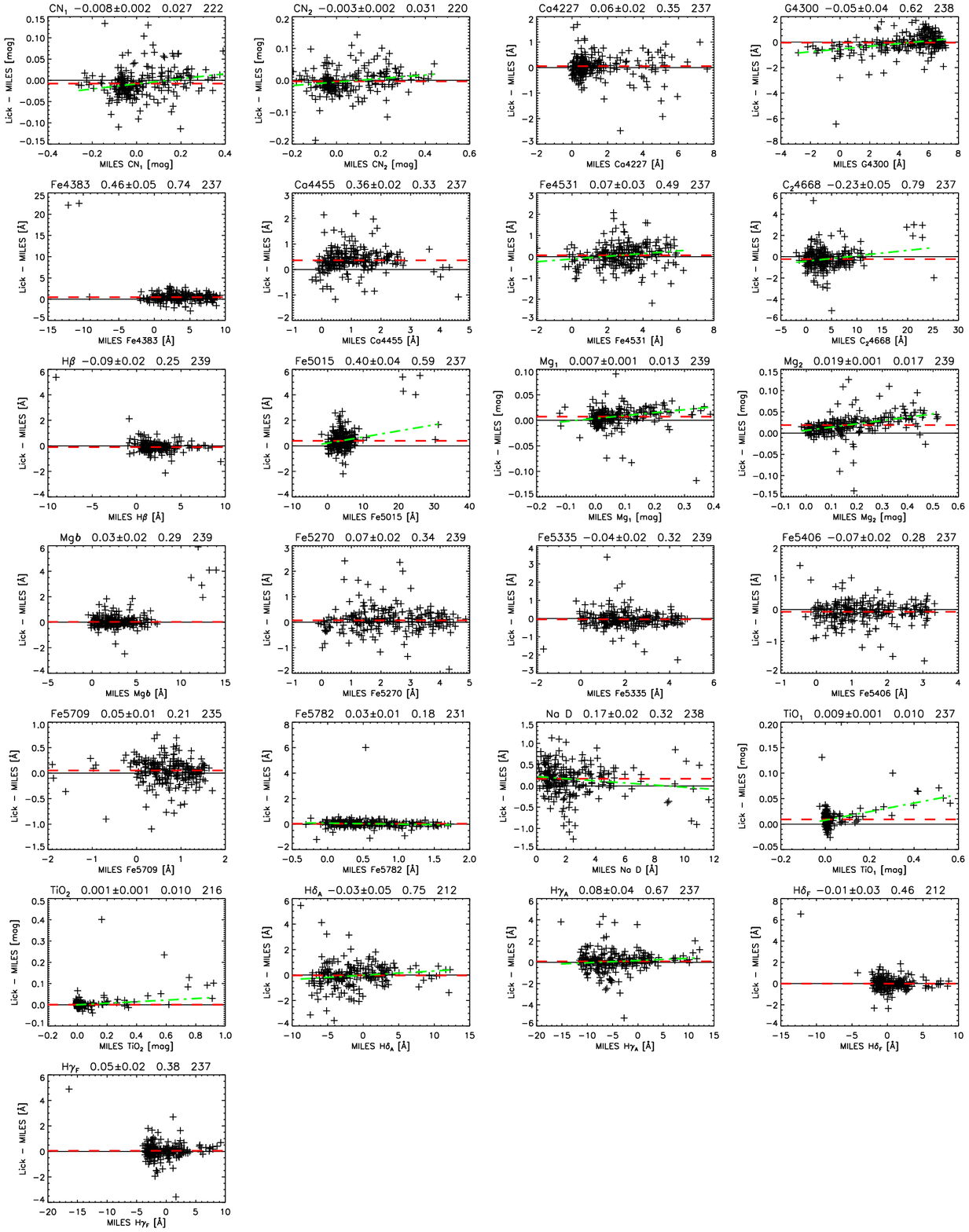} 
\label{loff}
\caption{Comparison of Lick/IDS index measurements with the MILES
  stellar library (Sanchez06) for stars in common (plus signs). The
  red dashed line in each panel shows the mean offset derived by a
  biweight estimator (see also Table~\ref{tab:offsets}). In the title
  of each panel we list the index name, the index offset and
  associated error, a robust estimate of the 1$\sigma$ scatter and the
  number of stars used in the comparison. A subset of the indices show
  evidence for  line-strength dependent offset. For indices where this
  is a significant effect (2$\sigma$) we also show a robust fit to the
  relation (green dash-dotted line). See text for details.} 
\label{fig:Milesoffset}
\end{figure*}

For this project and instrument configuration there were no stars
observed in common with the original Lick/IDS stellar library
\citep{Worthey94} which is the standard way to calibrate the index
measurements to the Lick system by deriving small offsets
\citep{Worthey97}. However, in this paper we made use of the MILES
stellar library \citep{Sanchez06} to determine the offsets for indices
measured on flux calibrated spectra to the original Lick/IDS system
assuming that the MILES library and our data are well
flux-calibrated. 

The spectra were first broadened to the Lick/IDS resolution with a
wavelength-dependent Gaussian assuming a spectral resolution of the
MILES library of 2.3 \AA\/ (FWHM). Then, we compared the index
measurements in common with the Lick stellar library for up to 239
stars depending on the availability of the measurements. The offsets
and associated errors (see Figure~\ref{fig:Milesoffset} and
Table~\ref{tab:offsets}) were derived with an outlier resistent
biweight estimator (IDL Astronomy library {\tt
  biweight\_mean.pro}). The offsets are generally small ($<$0.2 \AA)
but individual indices show larger offsets (e.g., Fe4383,
Fe5015). Additionally, several indices show evidence for line-strength
dependent offsets (e.g., CN$_2$, G4300, Mg$_2$; see also Vazdekis
1999). We used an outlier-resistant two-variable linear regression
method (IDL Astronomy library {\tt robust\_linefit.pro}) to
investigate the significance of such trends. For indices where we find
a slope different from zero with more than 2$\sigma$ significance we
show the fit in Figure~\ref{fig:Milesoffset} with a green dash-dotted
line and give the fit parameters in Table~\ref{tab:offsets}. 

We note that previous attempts to establish Lick offsets with the
\citealt{Jones97} stellar library
(\citealt{Norris06};\citealt{Worthey97}) have resulted in
significantly different offsets for some indices, perhaps most notably
the higher order Balmer lines. We ascribe the differences in offsets
to an imperfect flux calibration of the Jones library. Recently,
\citealt{Sanchez09} also used the MILES library to derive Lick offsets
and find similar results. However, significant differences remain for
individual indices (e.g., Mg\,$b$) where we favour our outlier robust
analysis.

For the present study, we apply the offsets listed in Table \ref{tab:offsets} (column 2). Only
one of the indices (Fe5015), used in this study to derive the stellar population
parameter estimates, shows evidence for a line-strength dependent offset.
Since the trend is very weak in the range of observed line-strength for this index
we apply only a normal offset.

\begin{table*} 
  \begin{minipage}{150mm} 
    \caption{Offsets to the Lick/IDS stellar library system derived from the MILES library.}
    \label{tab:offsets}
    \begin{tabular}{lcccrr} \hline
Index             & Offset           & rms   &  Number of stars & $A$  & $B$ \\
(1) & (2) & (3) & (4) & (5) & (6) \\
\hline
CN$_1$            & $-0.008\pm0.002$ & 0.027 &  222 & $-0.008\pm0.003$ & $ 0.060\pm0.014$ \\
CN$_2$            & $-0.003\pm0.002$ & 0.031 &  220 & $-0.005\pm0.004$ & $ 0.056\pm0.015$ \\
Ca4227            & $ 0.06\pm0.02$   & 0.35  &  237 & \multicolumn{2}{c}{-} \\
G4300             & $-0.05\pm0.04$   & 0.62  &  238 & $-0.545\pm0.050$ & $ 0.108\pm0.016$ \\
Fe4383            & $ 0.46\pm0.05$   & 0.74  &  237 & \multicolumn{2}{c}{-} \\
Ca4455            & $ 0.36\pm0.02$   & 0.33  &  237 & \multicolumn{2}{c}{-} \\
Fe4531            & $ 0.07\pm0.03$   & 0.49  &  237 & $-0.108\pm0.047$ & $ 0.064\pm0.021$ \\
C$_2$4668         & $-0.23\pm0.05$   & 0.79  &  237 & $-0.423\pm0.057$ & $ 0.051\pm0.013$ \\
H$\beta$          & $-0.09\pm0.02$   & 0.25  &  239 & \multicolumn{2}{c}{-} \\
Fe5015            & $ 0.40\pm0.04$   & 0.59  &  237 & $ 0.216\pm0.044$ & $ 0.048\pm0.011$ \\
Mg$_1$            & $ 0.007\pm0.001$ & 0.013 &  239 & $ 0.004\pm0.002$ & $ 0.057\pm0.009$ \\
Mg$_2$            & $ 0.019\pm0.001$ & 0.017 &  239 & $ 0.006\pm0.003$ & $ 0.078\pm0.006$ \\
Mg\,$b$           & $ 0.03\pm0.02$   & 0.29  &  239 & \multicolumn{2}{c}{-} \\
Fe5270            & $ 0.07\pm0.02$   & 0.34  &  239 & \multicolumn{2}{c}{-} \\
Fe5335            & $-0.04\pm0.02$   & 0.32  &  239 & \multicolumn{2}{c}{-} \\
Fe5406            & $-0.07\pm0.02$   & 0.28  &  237 & \multicolumn{2}{c}{-} \\
Fe5709            & $ 0.05\pm0.01$   & 0.21  &  235 & \multicolumn{2}{c}{-} \\
Fe5782            & $ 0.03\pm0.01$   & 0.18  &  231 & $ 0.077\pm0.022$ & $-0.076\pm0.025$ \\
Na~D              & $ 0.17\pm0.02$   & 0.32  &  238 & $ 0.219\pm0.026$ & $-0.025\pm0.010$ \\
TiO$_1$           & $ 0.009\pm0.001$ & 0.010 &  237 & $ 0.008\pm0.001$ & $ 0.084\pm0.008$ \\
TiO$_2$           & $ 0.001\pm0.001$ & 0.010 &  216 & $-0.000\pm0.002$ & $ 0.037\pm0.006$ \\
H$\delta_{\rmn A}$ & $-0.03\pm0.05$   & 0.75  &  212 & $-0.018\pm0.049$ & $ 0.033\pm0.012$ \\
H$\gamma_{\rmn A}$ & $ 0.08\pm0.04$   & 0.67  &  237 & $ 0.166\pm0.040$ & $ 0.021\pm0.008$ \\
H$\delta_{\rmn F}$ & $-0.01\pm0.03$   & 0.46  &  212 & \multicolumn{2}{c}{-} \\ 
H$\gamma_{\rmn F}$ & $ 0.05\pm0.02$   & 0.38  &  237 & \multicolumn{2}{c}{-} \\ 
        \hline 
    \end{tabular} 

 \medskip Note. Column (1) gives the index name, while column (2) gives the mean offset (Lick - MILES) to the Lick/IDS system. Column (3) gives a robust estimate of the 1$\sigma$ scatter and column (4) shows th number of stars used in the comparison for each index. Columns (5) and (6) give the coefficents of a linear fit to (Lick - MILES) = $A + B \times$~Lick where a significant slope was found.
  \end{minipage}
\end{table*} 
\section{Lick/IDS INDICES} \label{ap2}
In Table \ref{idm} we list the measured indices and their errors,
which we used for the SSP parameter estimation of the galaxies of our sample. The indices are transformed to the Lick/IDS system 
using the offsets described in Appendix~\ref{ap1}.

\begin{table*}
 \centering
 \begin{minipage}{180mm}
\caption[idm]{Measured central line-strength indices corrected to the Lick/IDS system, measured in the central 4 arcsec with one sigma error}
\label{idm}
\begin{tabular}{cccccccccc}
\hline
Galaxy	&	H$\delta_F$			&	H$\gamma_F$			&	Fe4383			&	H$\beta$			&	Fe5015			&	Mg{\it b}			&	Fe5270			&	Fe5335			&	Fe5406			\\
\hline
b0308	&	2.29	$\pm$	0.28	&	1.66	$\pm$	0.24	&	3.46	$\pm$	0.56	&	2.82	$\pm$	0.26	&	4.78	$\pm$	0.57	&	2.29	$\pm$	0.28	&	2.20	$\pm$	0.31	&	1.93	$\pm$	0.35	&	1.07	$\pm$	0.27	\\
d0216	&	3.73	$\pm$	0.26	&	2.75	$\pm$	0.23	&	1.07	$\pm$	0.61	&	2.70	$\pm$	0.27	&	3.19	$\pm$	0.60	&	1.65	$\pm$	0.30	&	1.81	$\pm$	0.33	&	1.14	$\pm$	0.38	&	1.12	$\pm$	0.27	\\
d0389	&	1.57	$\pm$	0.30	&	0.09	$\pm$	0.27	&	4.38	$\pm$	0.56	&	2.52	$\pm$	0.25	&	4.61	$\pm$	0.57	&	2.32	$\pm$	0.27	&	2.53	$\pm$	0.30	&	2.03	$\pm$	0.34	&	0.99	$\pm$	0.26	\\
d0490	&	1.72	$\pm$	0.39	&	0.56	$\pm$	0.34	&	4.07	$\pm$	0.78	&	2.54	$\pm$	0.35	&	4.30	$\pm$	0.77	&	2.26	$\pm$	0.38	&	2.62	$\pm$	0.42	&	1.98	$\pm$	0.47	&	1.38	$\pm$	0.35	\\
d0856	&	1.39	$\pm$	0.25	&	0.57	$\pm$	0.21	&	4.26	$\pm$	0.47	&	2.28	$\pm$	0.22	&	4.36	$\pm$	0.48	&	2.35	$\pm$	0.23	&	2.42	$\pm$	0.26	&	1.90	$\pm$	0.29	&	1.27	$\pm$	0.23	\\
d0990	&	1.13	$\pm$	0.25	&	0.40	$\pm$	0.21	&	4.26	$\pm$	0.47	&	2.47	$\pm$	0.21	&	4.73	$\pm$	0.46	&	2.44	$\pm$	0.22	&	2.38	$\pm$	0.25	&	2.09	$\pm$	0.28	&	1.29	$\pm$	0.21	\\
d1304	&	2.66	$\pm$	0.31	&	1.60	$\pm$	0.27	&	2.68	$\pm$	0.64	&	2.56	$\pm$	0.28	&	3.66	$\pm$	0.65	&	1.68	$\pm$	0.31	&	1.89	$\pm$	0.35	&	1.64	$\pm$	0.39	&	1.02	$\pm$	0.31	\\
d2019	&	1.24	$\pm$	0.35	&	0.63	$\pm$	0.29	&	3.98	$\pm$	0.64	&	2.48	$\pm$	0.30	&	4.70	$\pm$	0.68	&	2.16	$\pm$	0.33	&	2.70	$\pm$	0.37	&	1.98	$\pm$	0.42	&	1.45	$\pm$	0.31	\\
n0545	&	1.14	$\pm$	0.33	&	0.61	$\pm$	0.29	&	2.44	$\pm$	0.65	&	1.87	$\pm$	0.30	&	3.74	$\pm$	0.66	&	1.43	$\pm$	0.32	&	1.48	$\pm$	0.36	&	1.03	$\pm$	0.40	&	0.83	$\pm$	0.31	\\
n0725	&	2.51	$\pm$	0.39	&	1.27	$\pm$	0.36	&	1.55	$\pm$	0.85	&	2.01	$\pm$	0.40	&	3.46	$\pm$	0.88	&	0.19	$\pm$	0.45 	&	1.74	$\pm$	0.48	&	1.37	$\pm$	0.55	&	0.75	$\pm$	0.40	\\
n0929	&	0.80	$\pm$	0.24	&	-0.59	$\pm$	0.21	&	5.38	$\pm$	0.43	&	2.06	$\pm$	0.20	&	5.44	$\pm$	0.43	&	3.08	$\pm$	0.21	&	2.74	$\pm$	0.23	&	2.42	$\pm$	0.26	&	1.29	$\pm$	0.20	\\
n1167	&	1.86	$\pm$	0.31	&	0.71	$\pm$	0.27	&	2.68	$\pm$	0.63	&	2.13	$\pm$	0.29	&	3.11	$\pm$	0.63	&	1.63	$\pm$	0.30	&	1.84	$\pm$	0.34	&	0.96	$\pm$	0.39	&	0.67	$\pm$	0.29	\\
n1185	&	1.16	$\pm$	0.32	&	0.27	$\pm$	0.29	&	3.29	$\pm$	0.62	&	2.00	$\pm$	0.29	&	4.33	$\pm$	0.67	&	1.63	$\pm$	0.32	&	1.80	$\pm$	0.36	&	1.96	$\pm$	0.40	&	0.89	$\pm$	0.30	\\
n1254	&	1.36	$\pm$	0.26	&	0.28	$\pm$	0.22	&	3.62	$\pm$	0.50	&	1.94	$\pm$	0.23	&	4.24	$\pm$	0.49	&	2.38	$\pm$	0.24	&	2.03	$\pm$	0.27	&	1.92	$\pm$	0.30	&	1.24	$\pm$	0.22	\\
n1261	&	1.50	$\pm$	0.21	&	0.66	$\pm$	0.18	&	3.99	$\pm$	0.38	&	2.45	$\pm$	0.18	&	4.48	$\pm$	0.39	&	2.61	$\pm$	0.19	&	2.63	$\pm$	0.21	&	2.15	$\pm$	0.24	&	1.37	$\pm$	0.18	\\
n1308	&	1.22	$\pm$	0.34	&	0.46	$\pm$	0.30	&	3.48	$\pm$	0.68	&	2.33	$\pm$	0.31	&	4.85	$\pm$	0.68	&	2.49	$\pm$	0.33	&	2.14	$\pm$	0.38	&	2.06	$\pm$	0.42	&	1.20	$\pm$	0.31	\\
n1333	&	2.58	$\pm$	0.34	&	1.33	$\pm$	0.30	&	2.51	$\pm$	0.73	&	2.24	$\pm$	0.33	&	3.14	$\pm$	0.74	&	1.48	$\pm$	0.36	&	1.23	$\pm$	0.41	&	0.85	$\pm$	0.46	&	0.69	$\pm$	0.34	\\
n1348	&	1.83	$\pm$	0.32	&	-0.08	$\pm$	0.29	&	3.70	$\pm$	0.65	&	1.58	$\pm$	0.30	&	3.31	$\pm$	0.67	&	2.98	$\pm$	0.31	&	1.98	$\pm$	0.35	&	1.58	$\pm$	0.40	&	0.98	$\pm$	0.30	\\
n1353	&	3.34	$\pm$	0.31	&	2.69	$\pm$	0.27	&	0.89	$\pm$	0.72	&	3.06	$\pm$	0.30	&	3.12	$\pm$	0.71	&	1.76	$\pm$	0.34	&	1.68	$\pm$	0.38	&	1.38	$\pm$	0.43	&	1.03	$\pm$	0.32	\\
n1355	&	1.98	$\pm$	0.39	&	0.97	$\pm$	0.34	&	3.50	$\pm$	0.80	&	2.28	$\pm$	0.36	&	4.41	$\pm$	0.82	&	1.69	$\pm$	0.40	&	2.44	$\pm$	0.45	&	1.46	$\pm$	0.51	&	1.34	$\pm$	0.37	\\
n1389	&	1.92	$\pm$	0.33	&	0.79	$\pm$	0.29	&	2.91	$\pm$	0.68	&	2.09	$\pm$	0.31	&	2.71	$\pm$	0.71	&	1.67	$\pm$	0.33	&	1.50	$\pm$	0.38	&	1.30	$\pm$	0.43	&	0.50	$\pm$	0.33	\\
n1407	&	1.46	$\pm$	0.30	&	0.31	$\pm$	0.27	&	3.27	$\pm$	0.59	&	1.88	$\pm$	0.27	&	3.90	$\pm$	0.60	&	2.36	$\pm$	0.29	&	1.64	$\pm$	0.33	&	1.55	$\pm$	0.37	&	1.10	$\pm$	0.28	\\
n1661	&	1.39	$\pm$	0.36	&	0.44	$\pm$	0.33	&	4.29	$\pm$	0.72	&	2.08	$\pm$	0.34	&	4.12	$\pm$	0.77	&	1.65	$\pm$	0.37	&	1.75	$\pm$	0.42	&	1.46	$\pm$	0.47	&	1.11	$\pm$	0.35	\\
n1826	&	1.12	$\pm$	0.31	&	0.26	$\pm$	0.26	&	3.25	$\pm$	0.59	&	2.01	$\pm$	0.27	&	4.40	$\pm$	0.59	&	1.93	$\pm$	0.29	&	2.31	$\pm$	0.32	&	2.00	$\pm$	0.36	&	1.26	$\pm$	0.27	\\
n1861	&	1.34	$\pm$	0.31	&	0.23	$\pm$	0.27	&	4.35	$\pm$	0.59	&	1.94	$\pm$	0.27	&	4.78	$\pm$	0.58	&	2.62	$\pm$	0.28	&	2.74	$\pm$	0.32	&	2.18	$\pm$	0.36	&	1.58	$\pm$	0.28	\\
n1945	&	2.13	$\pm$	0.31	&	1.17	$\pm$	0.28	&	4.12	$\pm$	0.62	&	2.59	$\pm$	0.29	&	3.20	$\pm$	0.67	&	2.23	$\pm$	0.31	&	1.92	$\pm$	0.36	&	1.37	$\pm$	0.40	&	1.20	$\pm$	0.30	\\
																																			
\hline
\end{tabular}
\end{minipage}
\end{table*}

\section{Comparison of the estimated SSP parameters} \label{ap3}
In this section we present several comparisons of the SSP parameters
derived from different SSP models, using different methods and also
different data sets. We also provide the derived age and metallicity
from the AV/MILES model for our data set in Table \ref{agmt_av}.  
\begin{table}
 \caption{ Derived stellar population parameters using the AV/MILES model at 11 \AA.}
 \label{agmt_av}
\begin{tabular}{lrl}
\hline  \smallskip
Galaxy	&	Age, Gyr 				&	   [Z/H], dex	\\
\hline
VCC	0308	&	1.8	$^{+	0.1	}_{	-0.3	}$&$	-0.01	\pm	0.17	$\\
\hline
VCC	0216	&	2.2	$^{+	0.0	}_{	-0.0	}$&$	-0.85	\pm	0.05	$\\
VCC	0389	&	3	$^{+	0.4	}_{	-0.2	}$&$	-0.16	\pm	0.09	$\\
VCC	0490	&	2.5	$^{+	0.3	}_{	-0.2	}$&$	-0.06	\pm	0.17	$\\
VCC	0856	&	2.8	$^{+	0.2	}_{	-0.0	}$&$	-0.16	\pm	0.09	$\\
VCC	0990	&	3	$^{+	0.2	}_{	-0.2	}$&$	-0.16	\pm	0.07	$\\
VCC	1304	&	2.7	$^{+	1.0	}_{	-0.3	}$&$	-0.68	\pm	0.14	$\\
VCC	2019	&	2.7	$^{+	0.2	}_{	-0.2	}$&$	-0.11	\pm	0.09	$\\
\hline
VCC	0545	&	15.6	$^{+	0.0	}_{	-5.1	}$&$	-1.06	\pm	0.12	$\\
VCC	0725	&	7.1	$^{+	2.1	}_{	-0.4	}$&$	-1.20	\pm	0.14	$\\
VCC	0929	&	3.7	$^{+	1.1	}_{	-0.5	}$&$	-0.01	\pm	0.09	$\\
VCC	1167	&	7.6	$^{+	9.1	}_{	-0.5	}$&$	-0.97	\pm	0.14	$\\
VCC	1185	&	7.1	$^{+	4.9	}_{	-0.9	}$&$	-0.70	\pm	0.19	$\\
VCC	1254	&	5.1	$^{+	0.7	}_{	-0.6	}$&$	-0.47	\pm	0.07	$\\
VCC	1261	&	2.7	$^{+	0.0	}_{	-0.2	}$&$	-0.09	\pm	0.10	$\\
VCC	1308	&	3.2	$^{+	1.3	}_{	-0.4	}$&$	-0.28	\pm	0.14	$\\
VCC	1333	&	6.2	$^{+	11.6	}_{	-1.1	}$&$	-1.06	\pm	0.24	$\\
VCC	1348	&	17.8	$^{+	0.0	}_{	-10.2	}$&$	-0.82	\pm	0.21	$\\
VCC	1353	&	2	$^{+	0.1	}_{	-0.1	}$&$	-0.68	\pm	0.14	$\\
VCC	1355	&	2.5	$^{+	0.5	}_{	-0.2	}$&$	-0.35	\pm	0.17	$\\
VCC	1389	&	16.7	$^{+	1.1	}_{	-9.6	}$&$	-1.13	\pm	0.19	$\\
VCC	1407	&	7.6	$^{+	1.1	}_{	-1.7	}$&$	-0.70	\pm	0.12	$\\
VCC	1661	&	6.6	$^{+	1.4	}_{	-2.2	}$&$	-0.68	\pm	0.22	$\\
VCC	1826	&	4.8	$^{+	0.7	}_{	-1.3	}$&$	-0.42	\pm	0.14	$\\
VCC	1861	&	3	$^{+	0.2	}_{	-0.4	}$&$	-0.09	\pm	0.12	$\\
VCC	1945	&	3	$^{+	1.2	}_{	-0.5	}$&$	-0.54	\pm	0.12	$\\

\hline
\end{tabular}
\end{table}

 \begin{figure}
\includegraphics[width=6.5cm]{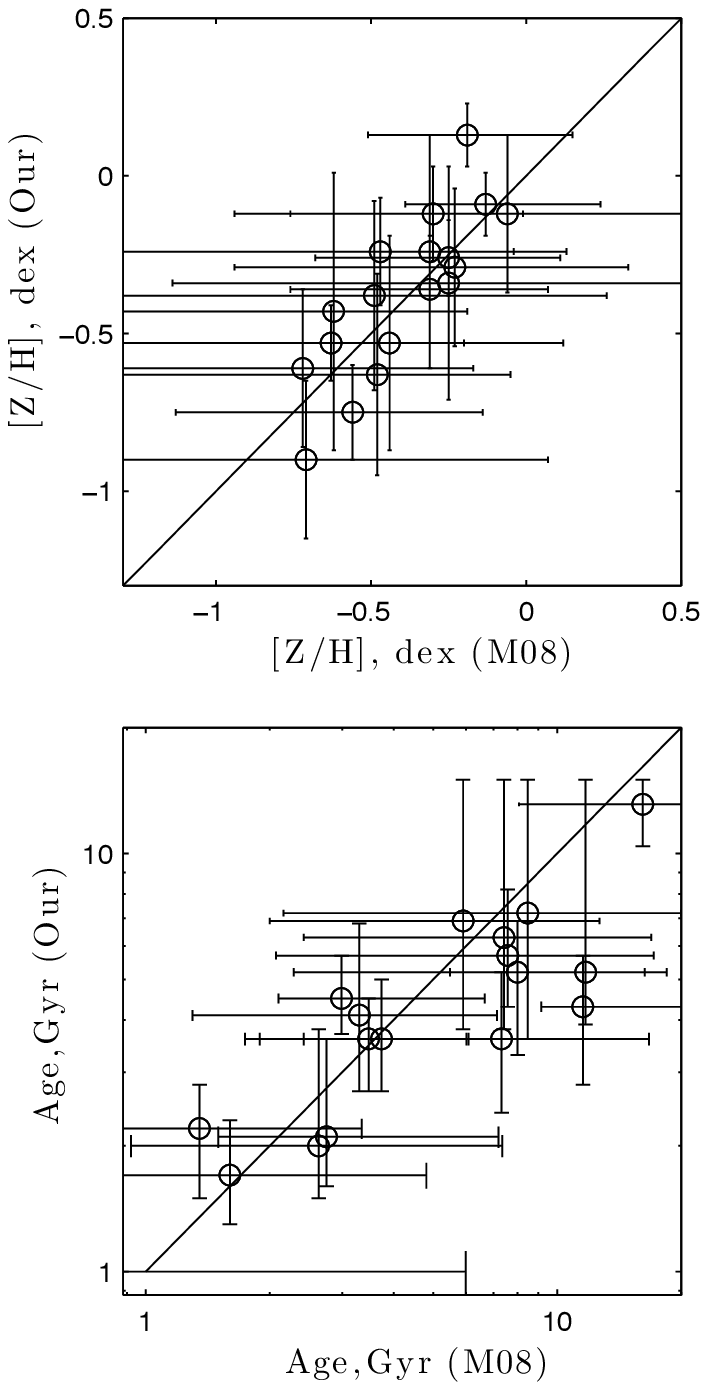}
\caption[fitid]{Comparison between the derived age (left) and
  metallicity (right) with
  TMB03 for our sample and that of M08. We use different approaches,
  i.e.\ following \citet{Proctor04} and \citet{Cardiel03} by us and by
  M08, respectively.}
\label{dms}
 \end{figure}

  \begin{figure*}
\includegraphics[width=170mm]{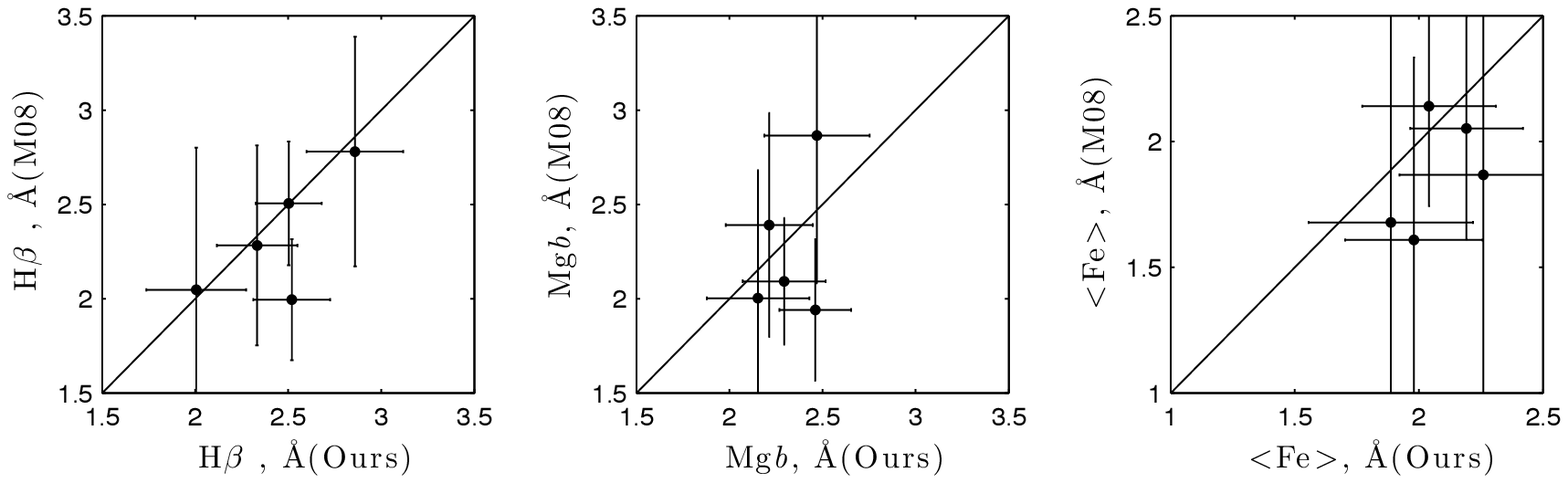}
\caption[fitid]{Comparison between the Lick indices measured from our
  and the M08 sample for the galaxies common to both samples, at the resolution 11 \AA} 
\label{lcc}
 \end{figure*}
 
   \begin{figure*}
\includegraphics[width=160mm]{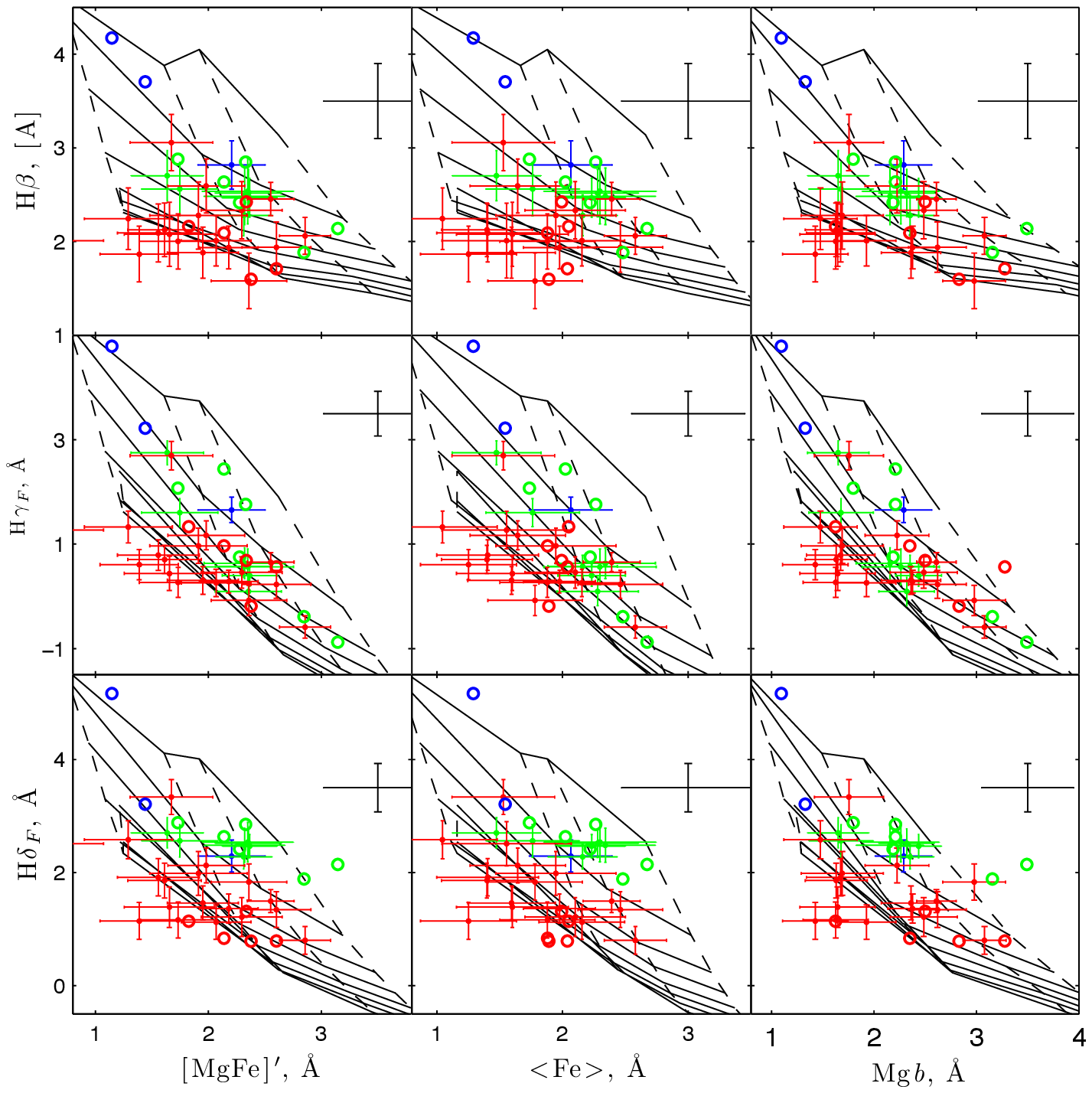}
\caption[fitid]{Lick/IDS index$-$index diagrams of the age-sensitive
  Balmer indices H$\beta$, H$\gamma_{F}$ and H$\delta_{F}$ versus the
  metallicity sensitive indices Mgb,$ <$Fe$>$ and [MgFe]$^\prime$, with
  overplotted model grids of TMB03 } 
\label{alind}
 \end{figure*}
 
  \begin{figure}
  \centering
\includegraphics[width=6.5cm]{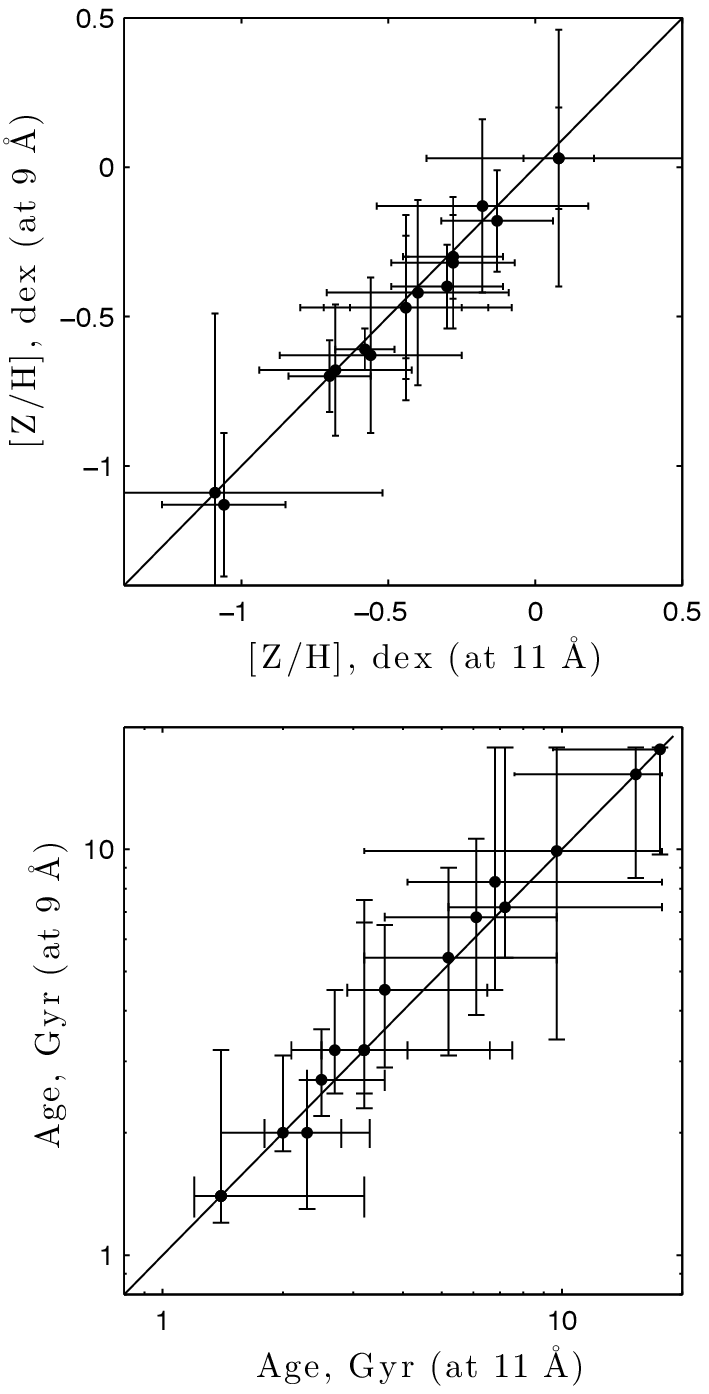}
\caption[cms_dms]{Comparison of derived stellar population parameters
  from the AV/MILES model at different resolutions for the M08 sample. } 
\label{cms_dms}
 \end{figure}

\label{lastpage}

\end{document}